\documentclass[preprint,aps,prd,superscriptaddress,floatfix,preprintnumbers,nofootinbib,altaffilletter]{revtex4-1}
\usepackage{graphicx} 
\usepackage{epstopdf}
\usepackage{slashed}
\usepackage{soul}
\usepackage{amsmath}
\usepackage{multirow}
\usepackage{tablefootnote}
\usepackage{comment}
\usepackage{color}
\usepackage{caption,subcaption}
\usepackage[normalem]{ulem}
\usepackage[T1]{fontenc}
\usepackage{graphicx}
\usepackage{mathrsfs}

\usepackage{amsmath}
\usepackage{amsfonts}
\usepackage{amssymb}
\usepackage{cleveref}
\usepackage{color}
\usepackage{bm}

\def\bea{\begin{eqnarray}}
\def\eea{\end{eqnarray}}
\def\beq{\begin{equation}}
\def\eeq{\end{equation}}
\def\be{\begin{equation}}
\def\ee{\end{equation}}
\def\nn{\nonumber}

\def\mC{\mathcal{C}}

\def\mI{\mathcal{I}}

\def\mL{\mathcal{L}}
\def\mM{\mathcal{M}}

\def\mO{\mathcal{O}}

\def\1{\textbf{1}}
\def\2{\textbf{2}}
\def\3{\textbf{3}}
\def\4{\textbf{4}}
\def\5{\textbf{5}}
\def\6{\textbf{6}}
\def\7{\textbf{7}}
\def\8{\textbf{8}}
\def\9{\textbf{9}}

\def\vep{{\varepsilon}}
\def\ep{{\epsilon}}
\def\lat{{\tilde{\lambda}}}
\def\ett{{\tilde{\eta}}}
\def\dota{{\dot{\alpha}}}
\def\<{\langle}
\def\>{\rangle}
\def\txtf{{\text{f}}}
\def\txtc{{\text{c}}}
\def\feri{{\mathsf{i}}}

\def\ferj{{\mathsf{j}}}

\def\non{\nonumber}

\def\bea{\begin{eqnarray}}
\def\eea{\end{eqnarray}}
\def\beq{\begin{equation}}
\def\eeq{\end{equation}}
\def\bsp{\begin{split}}
\def\esp{\end{split}}
\def\nn{\nonumber}

\def\lag{\langle}
\def\rag{\rangle}
\def\<{\langle}
\def\>{\rangle}
\def\vep{\varepsilon}

\def\larray{\left(\begin{array}{cccccccccccc} }
\def\rarray{\end{array} \right)}
\def\ddu{$\ddot{\text u}$}

\def\U{\text{U}}
\def\SU{\text{SU}}
\def\SO{\text{SO}}
\def\SL{\text{SL}}
\def\ISO{\text{ISO}}

\newcommand{\lrag}[1]{\langle#1\rangle}

\newcommand{\amp}[4]{($#1${}$#2${}$#3${}$#4$)}

\begin{document}

\title{Gauge invariance from on-shell massive amplitudes and tree unitarity}

\author{Da Liu}
\email{daeliu@ucdavis.edu}
\affiliation{Center for Quantum Mathematics and Physics (QMAP),  University of California, Davis, CA 95616, USA}

\author{Zhewei Yin}
\email{zhewei.yin@physics.uu.se}
\affiliation{Department of Physics and Astronomy, Uppsala University, 75108 Uppsala, Sweden}

\begin{abstract}

We study the three-particle and four-particle scattering amplitudes for an arbitrary, finite number of massive scalars, spinors and vectors by employing the on-shell massive spinor formalism.  We consider the most general three-particle amplitudes with energy growing behavior at most of $\mO(E)$. This is the special case of 
 the requirement of tree unitarity, which states that the $N$-particle scattering amplitudes at tree-level should grow at most as $\mO(E^{4-N})$ in the high energy hard scattering limit, i.e. at fixed non-zero angles. Then the factorizable parts of the four-particle amplitudes are calculated by gluing the on-shell three-particle amplitudes together and utilizing the fact that tree-level amplitudes have only simple poles. The contact parts of the four-particle amplitudes are further determined by tree unitarity, which also puts strong constraints on the possible allowed  three-particle coupling constants and the masses.  The derived relations among them converge to the predictions of gauge invariance in the UV theory. This provides a purely on-shell understanding of spontaneously broken gauge theories.

\end{abstract}

\preprint{UUITP-23/22}

\maketitle
\noindent

\tableofcontents

\section{Introduction}

In Ref. \cite{Weinberg:1995mt}, Weinberg took the point of view that the quantum field theory (QFT) is an inevitable outcome of the physical principles of quantum mechanics and special relativity. Starting from Wigner's definition of particles as irreducible representations of the inhomogeneous Lorentz group and by exploring the symmetries of the $S$-matrix, especially Lorentz invariance (covariance), as well as the cluster decomposition principle, it is possible to show that field theory is  a natural framework to describe physics at sufficiently low energy. The central role  in Wigner's classification is played by the little group for a given momentum,  which is defined as the subgroup of the Lorentz group that leaves the momentum unchanged. The general unitary Lorentz transformation on the Hilbert space, $U(\Lambda)$, can be induced by transformations  of the little group $W(\Lambda, p)$.

The $S$-matrix element is then given by the transition amplitude between the ``in'' and ``out'' states, which transforms as the direct product of one-particle states. The Lorentz covariance of $S$-matrix requires that there should exist one unitary operator, acting on both ``in'' and ``out'' states, which further leads to the commutation of the $S$-operator with the free Lorentz generators. Thus, one can show that  in the time-dependent perturbation theory, if the interaction operator can be written as an integral of a scalar density, which commutes at space-like or light-like separations: 
\beq
\label{eq:inter}
V(t) = \int d^3x \mathcal{V}(\mathbf x, t), \qquad \text{and}\qquad [\mathcal{V}(x), \mathcal{V}(x^\prime)] = 0 \qquad \text{for} \qquad (x - x^\prime)^2 \leq 0,
\eeq
then the $S$-matrix is Lorentz covariant. Furthermore, the cluster decomposition principle requires that the $S$-matrix factorizes for multi-particle processes which are sufficiently separated in space. It can be shown that if the Hamiltonian can be expressed as the sum of the products of annihilation and creation operators with coefficient functions only containing single three-momentum conservation delta function, the connected part of $S$-matrix will also only carry single momentum conservation delta function. This ensures that the $S$-matrix in the coordinate space satisfies the cluster decomposition principle. Such considerations in addition to  the requirement of Eq.~(\ref{eq:inter}) naturally call for quantum fields as building blocks of $\mathcal{V}(x)$.

However, huge progress on studying the scattering amplitudes for gauge theory and gravity has been made in recent years (see Refs. \cite{Bern:2007dw,Britto:2010xq,Elvang:2013cua,Dixon:2013uaa,Henn:2014yza,Cheung:2017pzi,Travaglini:2022uwo} for nice reviews), suggesting that  pure on-shell ways to determine the $S$-matrix, without the notion of local quantum fields, are worth exploring.
The massless helicity amplitudes are natural functions of spinor-helicity variables \cite{Berends:1981rb,DeCausmaecker:1981jtq,Kleiss:1985yh,Gastmans:1990xh,Xu:1986xb,Gunion:1985vca}. At tree-level
they are rational functions of spinor products and only have simple poles. In particular, the marvellous simplicity of the maximal helicity violating (MHV) $n$-gluon amplitude (the amplitude with the maximal number of same helicities in the all-momenta-incoming convention) deduced by Refs. \cite{Parke:1986gb,Berends:1987me} has suggested that there may exist alternative ways to calculate the helicity amplitudes for the gauge theory other than the conventional field-theory Feynman-diagram approach.

Indeed,  Witten's formulation of perturbative gauge theory in the twistor space~\cite{Witten:2003nn} has motived the  Cachazo-Svrcek-Witten construction of the tree-level amplitudes using MHV diagrams \cite{Cachazo:2004kj} and finally, the discovery of on-shell Britto-Cachazo-Feng-Witten (BCFW) recursion relations \cite{Britto:2004ap,Britto:2005fq} has provided an efficient and elegant way to determine the tree-level amplitudes of Yang-Mills theory from their singularities.  Note that Weinberg's argument of the Lorentz-covariance of the $S$-matrix is ensured by the recursion relations. This has motived active research in the quest for the dual formulation of QFT, in which the symmetries and the simplicities of the amplitudes, such as  the infinite-dimensional Yangian symmetry of $\mathcal{N} = 4$ planar supersymmetric Yang-Mills theory, are manifest, while the notion of locality or even space-time may not be apparent~\cite{Arkani-Hamed:2008bsc,Arkani-Hamed:2008owk,Arkani-Hamed:2009ljj,Arkani-Hamed:2010zjl,Arkani-Hamed:2012zlh}. The formulation of  planar $\mathcal{N} = 4$  SYM amplitudes  as the volume of ``Amplituhedron'' in Ref. \cite{Arkani-Hamed:2013jha} provides such an example: locality and unitary emerge from positivity geometry.   See also Refs.~\cite{Arkani-Hamed:2016rak,Rodina:2016mbk,Rodina:2016jyz,Rodina:2018pcb} for deriving locality and unitary from other principles, e.g. gauge invariance or infrared behavior.
Other interesting and exciting approaches include the color-kinematics duality and the double copy~\cite{Bern:2008qj,Bern:2010ue,Bern:2019prr}, the Cachazo-He-Yuan formalism based on scattering equations~\cite{Cachazo:2013gna,Cachazo:2013hca,Cachazo:2013iea,Cachazo:2014xea} etc.

From the more practical point of view, the idea of constructing amplitudes from the on-shell data has found many applications in the effective field theory (EFT), especially the EFT's that have enhanced soft limit, where on-shell constructability becomes possible~\cite{Cheung:2015ota,Cheung:2016drk,Cheung:2018oki,Elvang:2018dco,Low:2019ynd}. It has nicely explained the  one-loop non-renormalization patterns of  dimension-six operators in the Standard Model (SM) EFT~\cite{Cheung:2015aba}  and leads to new  non-renormalization theorem of operator mixing~\cite{Bern:2019wie}.  It also leads to the non-interference between the SM 4-point (-pt) amplitudes involving at least one transverse vector boson and the corresponding  linear dimension-six operator contributions~\cite{Azatov:2016sqh}. Calculations of anomalous dimensions of the effective operators have also been performed recently by using the on-shell amplitudes in Refs. \cite{Baratella:2020lzz,Baratella:2020dvw,Jiang:2020mhe, EliasMiro:2020tdv,Bern:2020ikv,EliasMiro:2021jgu}. Furthermore, the simplicity of on-shell helicity amplitudes can be used to enumerate the independent EFT operators, which was first demonstrated in the context of  a gauge singlet scalar or vector coupled to gluons in Ref.~\cite{Shadmi:2018xan} and further employed in Ref.~\cite{Ma:2019gtx,Durieux:2019siw,AccettulliHuber:2021uoa}.

Another step towards the on-shell formulation of QFT  has been put forward by Ref.~\cite{Arkani-Hamed:2017jhn}, in which the on-shell formalism for scattering amplitudes of general masses and spins has been systematically developed. The  massive particles carry $\SU (2)$ little group indices in the form of completely symmetric tensor representations, and as a result, the Lorentz-covariance of the scattering amplitudes for spin-$S$ particles has manifested itself as rank-2$S$ tensor. The  on-shell three-particle amplitudes can be constructed systematically by the use of massive spinor kinematic variables $\lambda^I_\alpha$, $\tilde{\lambda}^I_{\dot \alpha}$, and the four-particle scattering amplitudes can be derived by the fact that tree-level amplitudes have only simple poles and the residues are determined by unitary in the form of consistent factorization.

The presence of spurious ``non-local''  poles in the 3-pt on-shell massless helicity amplitudes (in the complex momenta scheme) and the requirement of consistent factorization for 4-pt amplitudes have put strong constraints on the possible structure of the interacting massless particles, such as the Yang-Mills structure for the multiple self-interacting massless spin-1 particles, and the universal couplings to the massless spin-2 particles \cite{Benincasa:2007xk}. In a similar fashion, one should be able to understand the structure of the spontaneously broken gauge theory as the consequence of the locality and perturbative unitarity constraints on the massive amplitudes. It is known that the on-shell 3-pt massive amplitudes can turn the spurious poles of massless amplitudes into some kind of mass singularities, and the Higgs mechanism can be understood as the infrared (IR) unification of different massless amplitudes in the ultraviolet (UV) \cite{Arkani-Hamed:2017jhn}. The $1/m$ mass singularities in general lead to energy growing behaviors for higher-pt amplitudes involving the longitudinal components of the massive gauge bosons.

Famously, it has been proved in the 1970's by the authors of Refs. \cite{Cornwall:1973tb,Cornwall:1974km,LlewellynSmith:1973yud} that any  tree unitary theory of massive vector bosons (with general interactions with scalars and spinors) is equivalent to a spontaneously broken gauge theory. Here tree unitarity means that the $N$-particle scattering amplitudes at tree level should scale at most  as $E^{4-N}$ in the high energy ($E$) limit at fixed, non-zero angles, i.e. the hard scattering limit. It can be argued that non-tree-unitary theories will not be renormalizable or in the modern effective field theory language, it will have very low cutoff.   In this paper, we aim to understand this from a completely on-shell point of view, using the aforementioned massive spinor helicity formalism. We will study the 3-pt and 4-pt scattering amplitudes for an arbitrary, finite spectrum of massive scalars, spinors and vectors, deriving the consequence of tree unitarity. Note that the general 3-pt on-shell massive amplitudes and 4-particle contact  terms for the SM and EFTs have been constructed by~\cite{Durieux:2019eor,Durieux:2020gip,Balkin:2021dko,Christensen:2018zcq,Li:2020tsi,Dong:2021yak,Dong:2022mcv,DeAngelis:2022qco}.  Ref.~\cite{Durieux:2019eor} has derived the constraints  among the  relevant coupling and mass parameters from perturbative unitary on the four-point $\psi^c \psi Zh$ amplitude. Similar work for the on-shell description of Higgs mechanism in the   SM electroweak sector has been studied in Ref. \cite{Bachu:2019ehv}. See Ref.~\cite{Cheung:2020tqz} for the consideration of  finite number of massless and massive scalar fields with arbitrary local interactions, where linearized symmetry and unification can emerge from soft theorems and perturbative unitarity. See also Refs.~\cite{Herderschee:2019dmc,Herderschee:2019ofc,Aoude:2019tzn,Falkowski:2020aso,Falkowski:2020fsu,Aoude:2020onz,Aoude:2020mlg,Low:2021ufv,Cohen:2022uuw,Abhishek:2022nqv} for related works.

 The paper is organized as follows. In Section~\ref{sec:review}, we first review the definition of the little group and its role in the classification of the irreducible representations of the inhomogeneous Lorentz group. Then we introduce on-shell massless and massive amplitudes and their interplay as the connection between UV and IR physics. In Section~\ref{sec:3pt}, we discuss the relations between polarization functions and massive spinor variables and  present the general relevant or marginal on-shell massive  3-particle amplitudes involving arbitrary numbers of  scalars, fermions and vector bosons. They are selected by studying their high energy limit and imposing the tree-level unitarity at 3-particle level, i.e. $\mM_3 \lesssim \mO (E)$. In Section~\ref{sec:4pt}, we move on to calculate the four-particle scattering amplitudes by obtaining the residues from gluing together the 3-particle on-shell massive amplitudes and then imposing the tree-level unitarity constraint, $\mM_4 \lesssim \mO (E^0)$, in the high energy limit. In addition to constraining possible contact parts of the 4-particle amplitudes,  we derive relations among the  coupling constants and show that they converge to gauge invariance in the UV theory, and a spontaneously broken symmetry in the IR. Section~\ref{sec:conclusion} contains our conclusion and outlook. Several appendices collect our conventions and useful formulae.

\section{The little group and the on-shell massless and massive amplitudes}
In this section, we review the basic concepts of the little group and illustrate how   the on-shell massless and massive amplitudes make the little group transformation manifest. The detailed discussion about the massless and massive spinor variables are presented in Appendix~\ref{app:mshv}  and Appendix~\ref{app:msv} respectively.
\label{sec:review}
\subsection{Review of the little group}
\begin{table}[th]
\begin{center}
\begin{tabular}{ |c|c|c| } 
 \hline
Standard momentum &Little group&  Standard Lorentz transformation\\ 
 \hline
$k^\mu = k(1,0,0,1)$& $\begin{array}{cc} \ISO (2) \\ J^2-K^1, \ -J^1 - K^2, \ J^3 \end{array}$ &$\begin{array}{ccc}L(p) = R(\hat{\mathbf p}) B(|\mathbf p|/k)\\
R(\hat{\mathbf p}) = e^{-i \phi J^3} e^{-i\theta J^2}\\
 B^0_{\ 0}(u)  =  B(u)^3_{\ 3} = \frac{u^2+1}{2 u}, \\
   B(u)^0_{\ 3}  =  B(u)^3_{\ 0} = \frac{u^2-1}{2 u} 
 \end{array}$ \\ 
   \hline
$k^\mu = m(1,0,0,0)$& $\begin{array}{cc} \SO (3) \\ J^1, \ J^2, \ J^3 \end{array}$& $\begin{array}{ccc}L(p) = R(\hat{\mathbf p}) B(|\mathbf p|)\\
R(\hat{\mathbf p}) = e^{-i \phi J^3} e^{-i\theta J^2}, \quad B(|\mathbf p|) = e^{- i \eta K^3}
 \end{array}$ \\ 
\hline
\end{tabular}
\end{center}
\caption{The little group for the massless and massive particles. The reference momenta and the corresponding standard Lorentz transformations are also shown. Here, $\hat{\mathbf{p}}$ stands for the unit vector along the direction of the 3-momentum, i.e. $\hat{\mathbf{p}} = (\sin\theta \cos\phi,\sin\theta \sin\phi,\cos\theta) $ and $\eta$ is the rapidity given by $\eta = \text{arctanh} \frac{|\mathbf p|}{E}$ . The explicit formulae for the Lorentz group generators in the vector representation are presented in Appendix~\ref{app:notation}.}
\label{tab:little}
\end{table}

 We start from the Wigner's definition of the little group~\cite{Wigner:1939cj}. In terms of Wigner's classification, the one-particle states can be defined as the irreducible representations of the inhomogeneous Lorentz group and the representations can be induced by the irreducible representations of the little group. Given a general momentum $p^\mu$, the little group is the subgroup of the homogeneous Lorentz group $\SO (3,1)$ or its universal covering group $\SL (2,\mC)$, which leaves the momenta of the particles the same. The classification can be performed using the reference momentum trick.
 
For massless particles, the reference momentum can be chosen as $k^\mu = k(1,0,0,1)$, where the little group associated with this reference momentum is simply the isometry group of the two-dimensional Euclidean space $\ISO (2)$.\footnote{Note that $\ISO (2)$ can be considered as the Innou-Wigner contraction of $\SO (3)$ with respect to its subgroup $\SO (2)$.} Actually, by using the explicit formulae in Eq.~(\ref{eq:lggenerators}), it is straightforward to show that the following combinations of the generators acting on the reference momentum will give zero four-vector:\footnote{Note that we have adopted the same convention for the definition of boosted generators as Peskin and Schroeder~\cite{Peskin:1995ev}, which is different from  that of Weinberg~\cite{Weinberg:1995mt} by a minus sign. See Appendix~\ref{app:notation} for details.}
\beq
J^2-K^1, \ -J^1 - K^2, \ J^3.
\eeq
To avoid the  continuum internal indices of the particles, only the subgroup $\SO (2) \simeq \U (1)$ is  considered. This means that the particles in the Hilbert space carry zero eigenvalues of the Hermitian operators corresponding to the first two generators. It is well-known that the representation is the helicity of the particles. The general momentum can be obtained by the standard Lorentz transformation, which can be chosen as\footnote{In this subsection, the boldface  letter $\mathbf p$ represents the three-momentum of the particle and in the following sections, we sometime use it for the massive on-shell spinors with symmetrized little group indices. As for the latter case, it is always associated with angle or square brackets, thus there should be no confusion.}
\beq
\label{eq:sdml}
L(p) = R(\hat{\mathbf p}) B(|\mathbf p|/k),
\eeq
where  the rotation $R(\hat{\mathbf p}) = \text{exp}(-i \phi J^3) \text{exp}(-i \theta J^2)$ transforms the $z$-axis into the direction of $\hat{\mathbf{p}} = (\sin\theta \cos\phi,\sin\theta \sin\phi,\cos\theta) $, and the boost $B(|\mathbf p|/k)$ is along the $z$-axis, with the non-zero components of $B$ given by
\beq
B^{0}_{\ 0} (u)= B^{3}_{\ 3}(u) = \frac{u^2+1}{2u},\quad  B^{0}_{\ 3}(u) = B^{3}_{\ 0} (u) = \frac{u^2-1}{2u}.
\eeq

For massive particles, we can choose the reference momentum as the momentum in the rest frame $k^\mu = m(1,0,0,0)$.  The little group is the rotation group $\SO (3)$ or its universal covering group $\SU (2)$. The generators are simply
\beq
J^1, \ J^2, \ J^3.
\eeq
The standard Lorentz transformation, which boosts the standard momentum $k^{\mu}$ to the general momentum $p^\mu$, can be chosen as
\beq
\label{eq:sdmv}
L(p) = R(\hat{\mathbf p}) B(|\mathbf p|) ,
\eeq
 where $B(|\mathbf p|) = \exp(-i \eta K^3)$ is the boost along the $z$-axis. Note that we have chosen a different standard Lorentz transformations from Ref. \cite{Weinberg:1995mt} and the reason will be clear later on. We summarize our previous discussions in Table~\ref{tab:little}. 

 In the Hilbert space, the state vector for the general momentum $p^\mu$ with helicity $\sigma$ of particle species $n$, $\Psi_{p,\sigma, n}$, can be obtained by 
 the unitary transformation $U(L(p))$ on the state vectors associated with the standard momentum $k^\mu$, $\Psi_{k,\sigma,n}$:
\beq
\Psi_{p,\sigma, n} = U(L(p)) \Psi_{k,\sigma,n}.
\eeq
It can be shown that once we normalize the states of the standard momentum:
\beq
(\Psi_{k^\prime,\sigma^\prime,n^\prime},\Psi_{k,\sigma,n}) =2 E_k\delta_{\sigma^\prime\sigma}  \delta_{nn^\prime}\delta^3(\mathbf k^\prime-\mathbf k) ,
\eeq
the states at general momenta have the following normalization:
\beq
(\Psi_{p^\prime,\sigma^\prime,n^\prime},\Psi_{p,\sigma,n}) = 2 E_p \delta_{\sigma^\prime\sigma}  \delta_{nn^\prime}\delta^3(\mathbf p^\prime-\mathbf p).
\eeq
Given the definition, under the  general proper orthochronous Lorentz transformation $\Lambda$, the state-vectors transform under the unitary operator  $U(\Lambda)$ as
\beq
U(\Lambda)  \Psi_{p,\sigma, n} = \sum_{\sigma^\prime} D_{\sigma^\prime \sigma}(W(\Lambda, p)) \Psi_{\Lambda p,\sigma^\prime, n}.
\eeq
 Here $W(\Lambda, p)$ is the little group element defined as
\beq
W(\Lambda, p) =  L^{-1}(\Lambda p) \Lambda L(p).
\eeq
For massive particles with spin $j$, $D_{\sigma^\prime \sigma}(W(\Lambda, p)) = D^{j}_{\sigma^\prime \sigma}(W(\Lambda, p))$ is an  irreducible unitary representation of $\SU (2)$ with dimension $2j +1$, while for massless particles, since helicity is a Lorentz invariant quantity, $D_{\sigma^\prime \sigma}(W(\Lambda, p))$ is diagonal with phase elements:
\beq
\label{eq:mslslg}
 D_{\sigma^\prime \sigma}(W(\Lambda,p)) = \text{exp}(-i \theta(\Lambda,p) \sigma) \delta_{\sigma^\prime \sigma}.
\eeq
It is important to point out for massive particles that when $\Lambda$ is the three-dimensional rotation $R$,  the little group rotation $W(\Lambda, p)$ remains the same as $R$,  i.e. $W(R, p) = R$, because $R$ is independent of the momentum $p$. This can be directly derived by using the explicit formula of the standard Lorentz transformation in Eq.~(\ref{eq:sdmv}).

The $S$-matrix elements are defined as the probability transition amplitudes from the in states $\Psi^+_{\alpha}$ to the out state  $\Psi^-_{\beta}$ as follows:
\beq 
S_{\beta\alpha} = (\Psi^-_\beta, \Psi^+_\alpha) ,
\eeq
with the state labels collectively given by  $\alpha = p_1\sigma_1n_1; p_2 \sigma_2 n_2;\cdots$, $ \beta = p_1^\prime \sigma_1^\prime  n_1^\prime; p_2^\prime \sigma_2^\prime n_2^\prime;\cdots$. The in and out states are transformed in the same way as the direct product of one-particle states. The Lorentz invariance of the $S$-matrix is defined as
\beq
S_{\beta\alpha} =  (U(\Lambda)\Psi^-_\beta, U(\Lambda)\Psi^+_\alpha) ,
\eeq
where the same unitary transformations acting on both in and out states are the essential part.
This will give us the Lorentz covariant property of the $S$-matrix:
\beq
S_{\beta \alpha}  =  \sum_{\bar{\sigma}_1,\bar{\sigma}_1^\prime, \cdots}D_{\bar\sigma_1\sigma_1}(W(\Lambda,p_1)) D_{\bar\sigma_2\sigma_2}(W(\Lambda,p_2)) \cdots D^*_{\bar\sigma_1^\prime\sigma_1^\prime}(W(\Lambda,p_1^\prime))   D_{\bar\sigma^\prime_2\sigma_2^\prime}^*(W(\Lambda,p_2^\prime)) \cdots S_{\Lambda \bar \beta, \Lambda \bar\alpha}.\label{eq:lrtcv}
\eeq
Here $\Lambda \bar\alpha$ stands for $ \Lambda p_1 \bar\sigma_1 n_1; \Lambda p_2 \bar\sigma_2 n_2;\cdots$ and the same applies to $\Lambda \bar\beta$. For massive particles with general spins, the Lorentz covariance tells us that the on-shell amplitudes are tensors under the little group $\SU (2)$, while for massless particles, the on-shell helicity amplitudes are subject to the $\U (1)$ little group phase transformations.   Since in our convention, we take all the momenta ingoing, the final particle states $\Psi_{p,\sigma,n}$ are represented by the analytical continuation $(-p, -\sigma, n)$.

\subsection{On-shell massless and massive amplitudes}
  The on-shell massless helicity amplitudes are naturally functions of spinor-helicity variables, which can be introduced by exploring the equivalence between the Lorentz group and the $\SL (2,\mC)$ group. The on-shell condition $p^2 = 0$ implies that the $2\times 2$ matrix  $p_{\alpha \dot \alpha} =  p_\mu \sigma^\mu_{\alpha \dot \alpha}$ has rank 1 and can be written as the direct product of two spinor vectors
(see Appendix~\ref{app:notation} for the summary of the notations and conventions):
\beq
\label{eq:spinorhelicity}
p_{\alpha\dot{\alpha}} =   \lambda_\alpha \tilde{\lambda}_{\dot{\alpha}} .
\eeq
Similar to the previous discussions about the induced general Lorentz transformation from the little group transformation, we can also specify the standard Lorentz transformation on the spinor variables~\cite{Arkani-Hamed:2017jhn}:
\beq
\lambda_\alpha(p) = D(L(p))_\alpha^{\ \beta}\lambda_\beta(k),
\eeq
and therefore, under the general Lorentz transformation $\Lambda$, the spinor variable has the  following little group transformation:
\beq
D(\Lambda) \lambda(p)  = D(W(\Lambda,p)) \lambda(\Lambda p).
\eeq
Here for real momenta, $D(W(\Lambda,p)) $ is just the $\U (1)$ little group phase factor $ e^{\frac i2\theta(\Lambda,p)}$, which corresponds to  $\sigma = -\frac 12$ in Eq.~(\ref{eq:mslslg}). For general complex momenta,  $D(W(\Lambda,p))$ will be a complex number $w\in \mC$, as the complexification of the little group $\U (1)$ is $\text{GL} (1,\mC)$.  In these definitions (conventions), the spinor $\lambda(\tilde \lambda)$ has helicity weight $-(+)\frac12$.  If we consider the helicity amplitudes  as functions of the spinor-helicity variables $\lambda,\tilde \lambda$, the Lorentz  covariance of the $S$-matrix in Eq.~(\ref{eq:lrtcv}) can be stated as follows:
\beq
\mM^{h_1,\cdots, h_n}(w_1\lambda_1,w_1^{-1}\tilde  \lambda_1; \cdots; w_n\lambda_n, w_n^{-1}\tilde{\lambda}_n ) = w_1^{-2h_1}\cdots w_n^{-2h_n}\mM^{h_1,\cdots, h_n}(\lambda_1,\tilde  \lambda_1; \cdots; \lambda_n, \tilde{\lambda}_n ),
\eeq
where for real momenta, $w_i = e^{\frac i2\theta(\Lambda,p_i)}$.

The beauty and power about the massless on-shell amplitudes are manifest from the fact that the 3-pt on-shell amplitudes in the scheme of complex momenta are uniquely fixed by the requirement of on-shell conditions, momentum conservation and the good behaviors under the real momentum limit. To be specific, the on-shell three-particle helicity amplitudes are given by
\beq
\label{eq:3ptham}
\begin{split} 
\mathcal{M}_3^{h_1,h_2,h_3} &= \left\{ 
\begin{array}{ll} 
 \left<12\right>^{h_3  - h_1 - h_2}  \left<23\right>^{h_1 - h_2 -h_3} \left<31\right>^{h_2 - h_3 - h_1}, \qquad  h_1 + h_2 + h_3 < 0\\
 \left[12\right]^{h_1+h_2 -h_3}  \left[23\right]^{h_2 + h_3 - h_1} \left[31\right]^{h_3 + h_1 - h_2}, \qquad  h_1 + h_2 + h_3 > 0\\
\end{array}
\right.
\end{split},
\eeq
which have smooth limits when we take the momenta as real quantity, i.e. $\lambda_1\propto \lambda_2 \propto\lambda_3$, or $ \tilde\lambda_1 \propto \tilde\lambda_2 \propto\tilde\lambda_3$ (see Appendix~\ref{app:mshv} for detail).  Note that under the parity transformation, the spinor-helicity variables transform as\footnote{We can check explicitly that under this transformation, $p_{\alpha\dot\alpha}$ changes to $p^{\dot\alpha \alpha}$ which is consistent with the parity transformation on the momentum $(p^0, \vec p) \rightarrow (p^0, -\vec p)$. The presence of the factor of $i$ is also consistent with the reality condition for the positive energy $\tilde{\lambda}_{\dot\alpha} = \lambda^*_\alpha$. One can also check this explicitly by using the explicit formulae of the spinor-helicity variables as functions of $(\theta, \phi)$ and noticing that under parity transformation, $\theta\rightarrow \pi - \theta$, $\phi \rightarrow \phi + \pi$.}
\beq
\lambda_\alpha \rightarrow i \tilde \lambda^{\dot \alpha}, \qquad \tilde{\lambda}^{\dot\alpha} \rightarrow i \lambda_\alpha,
\eeq
which results in the interchange between the angular and square brackets:
\beq
 \left<12\right> \leftrightarrow [12].
\eeq
Since the helicities change sign under the space inversion, the two cases in Eq.~(\ref{eq:3ptham}) are related by the parity transformation. 

A special case corresponds to the total helicity of $\pm 1$:
\beq
|h_1 + h_2 + h_3|=1,
\eeq
where by dimensional analysis, the coupling constants associated with the helicity amplitudes have mass dimension-zero. This corresponds to the marginal interaction terms in the classification of Wilson~\cite{Wilson:1971bg,Wilson:1973jj}. Let's focus on the case of $h_1 + h_2 + h_3 = 1$, then we have
\beq
\mathcal{M}_3^{h_1,h_2,h_3}  = \left[12\right]^{1 -2h_3}  \left[23\right]^{1 - 2h_1} \left[31\right]^{1 - 2h_2}.
\eeq
It can be immediately seen that there are always spurious poles for  amplitudes involving particles with helicities greater than or equal to one. When we try to calculate the residues of four-particle scattering amplitudes in one particular channel by gluing the 3-particle amplitudes together, they will always lead to poles in other channels. This plus the requirement of the unitarity in the form of consistent factorization have put strong constraints on  the allowed possible interaction types and coupling structures of 3-particle on-shell amplitudes~\cite{Benincasa:2007xk}. In particular, the self-interacting multiple spin-1 particles must have Yang-Mills structure, and the interactions between  fermions and the vector bosons must form a representation of the Lie algebra of the vector bosons. As we will see in the following discussion, in the case of massive vector bosons and fermions,  the same conclusion holds and the requirement of consistent factorization corresponds to imposing tree-level unitarity.

The massless spinor-helicity variables have been generalized to general masses and spins by Ref. \cite{Arkani-Hamed:2017jhn}.  In contrast to the spinor-helicity variables, the massive spinor variables carry little group $\SU (2)$ indices:
\beq
\lambda_\alpha \rightarrow \lambda_\alpha^I, \qquad \tilde{\lambda}_{\dot\alpha} \rightarrow \tilde{\lambda} ^I_{\dot\alpha},
\eeq
which corresponds to  the spin degrees of freedom of the particles. Similar to the massless case, one can obtain the massive spinor variables at general momentum $p^\mu$ from the standard Lorentz transformation of spinor variables at standard momentum $k^\mu = m(1,0,0,0)$:
\beq
\lambda^I_\alpha(p) = D(L(p))_\alpha^{\ \beta}\lambda^I_\beta(k),
\eeq
where $\lambda^{\frac12}_\alpha(k), \lambda^{-\frac12}_\alpha(k)$ correspond to spin-$z$ components of $+\frac12, -\frac12$ respectively.  Note that in our choice of standard Lorentz transformation in Eq.~(\ref{eq:sdmv}), $\lambda^{I}_\alpha(p)$ represents the spin component along the momentum-axis, i.e. the helicity. Once we specify the standard transformation, the general Lorentz transformation $\Lambda$ on the massive spinor variables are induced by the following little group transformation:
\beq
D(\Lambda) \lambda^I(p)  = D(W(\Lambda,p))^I_{\ J} \lambda^J(\Lambda p).
\eeq
Since the spin-$S$ particle carries $2S$ completely symmetric indices of $\SU (2)$, the Lorentz  covariance of the $S$-matrix in Eq.~(\ref{eq:lrtcv}) is equivalent to the statement that the corresponding scattering amplitudes are fully symmetric rank-$2S$ tensors of the massive spinor variables $\lambda^I, \tilde{\lambda}^I$. The momentum of the particle transforms trivially under the little group, thus can be constructed as an ``inner'' product of $\lambda^I, \tilde{\lambda}^I$:
\beq
\label{eq:massivespinor}
p_{\alpha\dot{\alpha}} = \varepsilon_{IJ}  \lambda^I_\alpha \tilde{\lambda}^J_{\dot{\alpha}}
= \lambda^I_\alpha \tilde{\lambda}_{I\dot{\alpha}},
\eeq
which can also be thought of as the sum of two rank 1 matrices.
As in the massless case, $\lambda^I_\alpha$ is independent of $\tilde{\lambda}_I$ for the general complex momenta and the limit of real momenta can be obtained by taking
\beq
\tilde{\lambda}_{I\dot\alpha} =\pm (\lambda^{I}_\alpha)^*,
\eeq
where the $+(-)$ sign depends on the energy being positive or negative, respectively. Note that another advantage for the massive spinor variables is the simple relations with the massless spinor variables as the high energy limit. To see this, we can always expand the spinors in the bases of the little group space as
\beq
\label{eq:msexp}
\lambda^{I}_\alpha = \lambda_\alpha \zeta^{-I} + \eta_\alpha \zeta^{+I}, \quad \tilde{\lambda}^I_{\dot{\alpha}} = \tilde{\lambda}_{\dot{\alpha}} \zeta^{+I} + \tilde{\eta}_{\dot{\alpha}} \zeta^{-I}.
\eeq
 In terms of the expansion above, the momentum matrix can be rewritten  as
\beq
p_{\alpha \dot{\alpha}}= \lambda_\alpha \tilde{\lambda}_{\dot{\alpha}} - \eta_\alpha \tilde{\eta}_{\dot{\alpha}}.
\eeq
As discussed in Appendix~\ref{app:msv}, with suitable sign convention, the on-shell condition of the momentum becomes
\beq
 \left<\lambda\eta\right> = m, \qquad [\tilde{\lambda}\tilde{\eta}] = m.
\eeq
We will choose $\lambda, \tilde{\lambda}$ as the surviving parts in the high energy limit, which will scale like $\sqrt{E}$. On the other hand, the sub-leading spinor variables  $\eta,\tilde{\eta}$ scale like $\frac{m}{\sqrt{E}}$. This doesn't mean that $\eta,\tilde{\eta}$ are totally irrelevant in the high energy limit; actually, there are always mass singularities associated with massive vector bosons. The relation between UV-massless on-shell amplitude and IR-massive amplitudes can be described as ``un-bold'' to ``bold'', with the subtlety for the massive spin-one or higher spin particles.  For example, for fermion-fermion-scalar amplitudes, we have:
\beq
 \<12\>   \leftrightarrow \<\1\2\>, \qquad  [12] \leftrightarrow [\1\2] 
\eeq
while for vector-vector-scalar amplitudes, we have:
\beq
\left(\frac{[12][23]}{[31]},  \frac{\lag12\rag\lag23\rag}{\lag31\rag}\right) \leftrightarrow \sqrt{2}\frac{[\1\2]\<\1\2\>}{m_2}
\eeq
Here the bold notation means that the little group indices  are completely symmetrized with appropriate Clebsch-Gordan coefficients~\cite{Arkani-Hamed:2017jhn,Durieux:2019eor}.  To be more explicit, we have:
\beq
[\1\2]\<\1\2\> \equiv  \frac{1}{\sqrt{2}}\left([1^{I_1}\2] \<1^{I_2} \2\> +  [1^{I_2}\2] \<1^{I_1} \2\> \right), \qquad I_1 \neq I_2
\eeq
In the latter case, we can choose the suitable $\eta,\tilde{\eta}$ as the reference spinor in order to IR-deform the massless 3-pt amplitudes (see Appendix~\ref{app:IRdef} for detail).

\section{General relevant and marginal 3-particle amplitudes}
\label{sec:3pt}
In this section, we will present the general on-shell massive 3-particle amplitudes relevant in our calculation. We only consider the particles with spin less than or equal to one and leave other cases for future possible work. We adopt a bottom-up approach and allow the coupling constants to be arbitrary and eventually we will see that the group structure and gauge invariance will emerge from the requirement of tree-level unitarity, i.e. $\mM_n \lesssim \mO ( E^{4-n} )$. In the same spirit of Ref.~\cite{Cornwall:1974km}, we consider arbitrary finite number of scalars, fermions  and vectors, by which we mean the Hilbert space consists of one-particle states labelled by their momenta, helicities and species $\Psi_{p,\sigma, n}$. Note that Ref. \cite{Durieux:2019eor} has studied the on-shell  3-pt massive  amplitude bases with the particle spectrum of the electroweak sector in SM in addition to one generation of fermions. We start from the discussion of the polarization functions and their relations with massless and massive spinor variables.

\subsection{Polarization functions}
\label{sec:polfun}
\begin{table}[th]
\small
\begin{center}
\begin{tabular}{ |c|c|c|c|c| } 
 \hline
Standard momentum & Polarization functions& Spinor variables \\ 
 \hline
$k^\mu = k(1,0,0,1)$&$\begin{array}{ccc} 
\epsilon^{+\mu} =\frac{1}{\sqrt{2}} \left(\begin{array}{c}
0\\
1\\
i\\
0
\end{array}\right),\quad
\epsilon^{-\mu} =\frac{1}{\sqrt{2}}\left(\begin{array}{c}
0\\
1\\
-i\\
0
\end{array}\right)
 \end{array}$ &$\begin{array}{ccc} 
 \lambda_\alpha = \sqrt{2k} \left(\begin{array}{c}
0\\
1
\end{array}\right),\\
 \tilde  \lambda^{\dot \alpha} = \sqrt{2k} \left(\begin{array}{c}
1\\
0
\end{array}\right),\\
\epsilon^-_{\alpha\dot{\alpha}} =\sqrt{2}  \frac{\lambda_\alpha \tilde\mu_{\dot{\alpha}}}{[\tilde\lambda\tilde \mu]}, \\
\epsilon^+_{\alpha\dot{\alpha}} = \sqrt{2} \frac{\mu_\alpha \tilde\lambda_{\dot{\alpha}}}{\lag\mu\lambda\rag}
 \end{array}$\\ 
   \hline
$k^\mu = m(1,0,0,0)$&$\begin{array}{ccc} 
\epsilon^{0\mu} = \left(\begin{array}{c}
0\\
0\\
0\\
1
\end{array}\right),
\epsilon^{+\mu} =\frac{1}{\sqrt{2}} \left(\begin{array}{c}
0\\
1\\
i\\
0
\end{array}\right),\
\epsilon^{-\mu} =\frac{1}{\sqrt{2}}\left(\begin{array}{c}
0\\
1\\
-i\\
0
\end{array}\right)\\
u^{+\frac12} = \sqrt{m} \left(\begin{array}{c}
1\\
0\\
1\\
0
\end{array}\right), \quad u^{-\frac12} =  \sqrt{m}\left(\begin{array}{c}
0\\
1\\
0\\
1
\end{array}\right), \\
v^{+\frac12} =  \sqrt{m}\left(\begin{array}{c}
1\\
0\\
-1\\
0
\end{array}\right), \quad 
v^{-\frac12} =  \sqrt{m}\left(\begin{array}{c}
0\\
1\\
0\\
-1
\end{array}\right)
 \end{array}$  & $\begin{array}{cc}\lambda^I_{\alpha}=\sqrt{m}\,\delta^I_\alpha, \\
  \tilde{\lambda}^{I \dot\alpha} =  \sqrt{m}\, \delta^{I \dot\alpha}\\ \\
\epsilon^{IJ}_{\alpha\dot\alpha} = \frac{\sqrt{2}}{m}  \lambda_\alpha^{\{I_1}   \tilde\lambda_{\dot{\alpha}}^{I_2\}} \\  \\
u^I=\left( \begin{array}{ccccc}
 \lambda^I_\alpha \\
 \tilde{\lambda}^{I\dot \alpha}
 \end{array}\right)\\
  v^I=\left( \begin{array}{ccccc}
\lambda^I_\alpha \\
 -\tilde{\lambda}^{I\dot \alpha}
 \end{array}\right)\\
 \bar{u}_I = (-\lambda_I^\alpha,\tilde{\lambda}_{I\dot\alpha}), \\  \bar{v}_I = (\lambda_I^\alpha,\tilde{\lambda}_{I\dot\alpha})
   \end{array}$ \\ 
\hline
$\begin{array}{c}
\text{General momentum:}\\
p^\mu = (E, \mathbf p)
\end{array}$
& $\begin{array}{ccc}\epsilon^\mu (p) =  L^\mu_{\ \nu}(p) \epsilon^\nu(k),\\
u^a(p) = D(L(p))^a_{\ b} u^b(k), \\ v^a(p) = D(L(p))^a_{\ b} v^b(k)
\end{array}$   & $\begin{array}{ccc}\lambda^I_{\alpha}(p) = D(L(p))_{\alpha }^{\ \beta}\lambda^I_{\beta}(k)\\
\tilde\lambda_{I\dot\alpha}(p) = D^*(L(p))_{\dot\alpha}^{\ \dot \beta}\tilde\lambda_{I\dot\beta}(k)\\
\lambda^{\frac12}_{\alpha} =    \sqrt{E - p} \left(\begin{array}{c}
 \cos\frac{\theta}{2}e^{-i\frac{\phi}{2}}\\
 \sin\frac{\theta}{2}e^{i\frac{\phi}{2}}\\
\end{array}\right)\\
 \lambda^{-\frac12}_{\alpha} =     \sqrt{E+p}\left(\begin{array}{c}
- \sin\frac{\theta}{2}e^{-i\frac{\phi}{2}}\\
\cos\frac{\theta}{2}e^{i\frac{\phi}{2}}\\
\end{array}\right)
\\
 \tilde{\lambda}_{I\dot\alpha} =( \lambda^I_\alpha)^*
\end{array}  $\\
\hline
\end{tabular}
\end{center}
\caption{Polarization functions and spinor variables at  standard and general momentum.  See the main text for detailed discussion. For the general momentum for the massless particles, $\lambda_\alpha$ coincides with  the high energy limit of $\lambda^{-\frac12}_\alpha$.}
\label{tab:wavefunction}
\end{table}

In Ref. \cite{Weinberg:1995mt}, the  polarization functions are obtained by requiring that the quantum fields constructed out of them and the annihilation (creation) operators transform  linearly under the Lorentz group and especially independent of the space-time coordinates. Specifically, the polarization functions satisfy the following conditions:
\beq
\label{eq:wfgm}
\begin{split}
\sum_{\bar \sigma} u_{\bar \ell}(\mathbf p_ \Lambda, \bar{\sigma},n) D_{\bar \sigma \sigma}^{(j_n)}(W(\Lambda,p)) =  \sum_{\ell} D_{\bar \ell \ell} (\Lambda)u_\ell(\mathbf p, \sigma, n),\\
\sum_{\bar \sigma} v_{\bar \ell}(\mathbf p_ \Lambda, \bar{\sigma},n) D_{\bar \sigma \sigma}^{(j_n)*}(W(\Lambda,p)) = \sum_{\ell} D_{\bar \ell \ell}(\Lambda) v_\ell(\mathbf p, \sigma, n),\\
\end{split}
\eeq
where $u_\ell$ and $v_\ell$ are the polarization functions associated with annihilation and creation operator respectively and $D_{\bar\ell\ell}(\Lambda)$ belongs to any irreducible representation of the Lorentz group.

The consequences of the above formulae can be explored by the special cases. For $\mathbf p = 0$ and $\Lambda = L(q)$ such that $W(\Lambda,p) = 1$, we have the following useful identities:
\beq
\begin{split}
u_{\bar \ell}(\mathbf q, \sigma,n) = \sum_{\ell} D_{\bar \ell \ell} (L(q))u_\ell(0, \sigma, n),\\
 v_{\bar \ell}(\mathbf q, \sigma,n)  = \sum_{\ell} D_{\bar \ell \ell}(L(q)) v_\ell(0, \sigma, n),\\
\end{split}
\eeq
which just tell us that the wave functions at general momentum can be obtained by the Lorentz transformations of the wave functions at zero momentum. To obtain the wave functions at zero momentum, we can take again $\mathbf p = 0$ but $\Lambda = R$, which is a three-dimensional rotation. This time, we have
\beq
\label{eq:wfzm}
\begin{split}
\sum_{\bar \sigma} u_{\bar \ell}(0, \bar{\sigma},n) D_{\bar \sigma \sigma}^{(j_n)}(R) =\sum_{\ell} D_{\bar \ell \ell} (R)u_\ell(0, \sigma, n),\\
\sum_{\bar \sigma} v_{\bar \ell}(0, \bar{\sigma},n) D_{\bar \sigma \sigma}^{(j_n)*}(R) =\sum_{\ell} D_{\bar \ell \ell} (R)v_\ell(0, \sigma, n).\\
\end{split} 
\eeq
This establishes the relations between the representations of the little group (spins or helicities) and the representations of the polarization functions under the Lorentz group (or more precisely, the rotation subgroup). The solutions of the above equation for spin-1 and spin-1/2 can be found by exploring the explicit formulae of the representation matrices and the results are shown in Table~\ref{tab:wavefunction}. 

To establish the relation between the polarization functions and the massive spinor variables, we first recall  the fact that the spin-$j$ representations of the rotation group can be treated as symmetrized direct products of $2j$ spin-1/2 representations. 
The normalized tensor state of $|j,\sigma\rangle$ corresponds to the following tensor components with $2j$ indices~\cite{Georgi:1999wka}:
\beq
\label{eq:norm}
 \left(\begin{array}{cc}
2 j\\
j+ \sigma
\end{array}\right)^{-1/2}v^{s_1\cdots s_{2j}}_{j,\sigma},
\eeq
where the completely symmetric tensor $v^{s_1\cdots s_{2j}}_{j,\sigma}$ is equal to one if there are $j+\sigma$  values of  spin $\frac 12$ and $j-\sigma$ values  of  spin $-\frac 12$ and zero otherwise. The normalization pre-factor comes from the fact that there are  $\left(\begin{array}{cc}
2 j\\
j+ \sigma
\end{array}\right)$ of possibilities. 
It also applies that the general irreducible representation of proper orthochronous  Lorentz group can be thought as direct sum of spins of two particles $(j_1,j_2)$, which are representations of complexified direct sum of two $\SU (2)$ Lie algebra $\mathfrak{su} (2) \oplus_\mC \mathfrak{su} (2)$:
\beq
\mathbf J_1 = \frac 12 (\mathbf J + i \mathbf K), \qquad \mathbf J_2 = \frac 12 (\mathbf J - i \mathbf K).
\eeq
In the formalism of completely symmetric tensor representations of $\SU (2)$,  we can think of the indices $\ell,\bar\ell$  in Eq.~(\ref{eq:wfzm}) as collections of $2j_1$ two-value indices $ \alpha_1\cdots \alpha_{2j_1}$ and $2j_2$ two-value indices  $ \dot{\alpha}_1\cdots \dot{\alpha}_{2j_2}$. The spin label $\sigma$ for particle of spin-$j_n$ can be treated as $2 j_n$ two-value indices $I_1,\cdots I_{2j_n}$. Actually, we only need the special case of $j_n = j_1 + j_2$. In this special case, the solutions to Eq.~(\ref{eq:wfzm}) can be obtained by the completely symmetric product of the building blocks $u_\alpha(0, I), u_{\dot\alpha}(I,0)$:
\beq
u_\alpha(0,I) = \sqrt{m} \delta_{I\alpha}, \qquad u_{\dot\alpha}(I,0) =\sqrt{m} \delta_{I\dot\alpha}.
\eeq
These are the massive spinor variables $\lambda,\tilde \lambda$ with the following index convention:
\beq
\lambda^I_{\alpha}(0) = u_\alpha(0,I) , \qquad \tilde{\lambda}^{I\dot \alpha}(0) = u_{\dot\alpha}(0,I) .
\eeq
The normalization is chosen such that:
\beq
\lambda^I_{\alpha}(0) \tilde{\lambda}_{I\dot \alpha}(0) = m \sigma^0_{\alpha\dot\alpha} = (p_\mu \sigma^\mu)_{\mathbf p =0},
\eeq
where $I,\dot\alpha$ are lowered by the antisymmetric tensor $\varepsilon_{IJ}$, $\varepsilon_{\dot\alpha\dot\beta}$.
The massive spinor variables at general momentum are given by the standard Lorentz transformation of the zero-momentum spinors:
\beq
\lambda^I_{\alpha}(\mathbf p) = (e^{-i\phi \frac{\sigma^3}{2}}e^{-i\theta \frac{\sigma^2}{2}}e^{-\eta \frac{\sigma^3}{2}})_{\alpha \alpha^\prime} \lambda^I_{\alpha^\prime}(0), \qquad \tilde \lambda^{I\dot\alpha}(\mathbf p) = (e^{-i\phi \frac{\sigma^3}{2}}e^{-i\theta \frac{\sigma^2}{2}}e^{\eta \frac{\sigma^3}{2}})^{\dot\alpha \dot\alpha^\prime} \tilde\lambda^{I\dot\alpha^\prime}(0),
\eeq
where  $\eta$ is the rapidity defined as $\cosh \eta = E/m$ and
as a consequence:
\beq
\lambda^I_{\alpha}(\mathbf p) \tilde{\lambda}_{I\dot \alpha}(\mathbf p) = p_\mu \sigma^\mu.
\eeq

We then make some comments on the massless particles.  For the scalar and spinor representations, satisfying Eq.~(\ref{eq:wfgm}) in the massless version is straightforward and there is no subtlety in taking the massless limit. For spin larger than or equal to one, it is not possible to satisfy the massless version of Eq.~(\ref{eq:wfgm}):
\beq
\label{eq:wfgm2}
\begin{split}
 u_{\bar \ell}(\mathbf p_ \Lambda, \bar{\sigma}) e^{-i\sigma\theta(p,\Lambda)} = \sum_{\ell} D_{\bar \ell \ell} (\Lambda)u_\ell(\mathbf p, \sigma)\\
\end{split}
\eeq
where $\theta(p,\Lambda)$ is the rotation angle of the little group transformation:
\beq
W (\Lambda, p) = L^{-1}(\Lambda p) \Lambda L(p) = S(\alpha(p,\Lambda),\beta(p,\Lambda)) R(\theta(p,\Lambda)).
\eeq
Here $ S(\alpha,\beta) $ is the invariant abelian subgroup of the little group $\ISO (2)$ and $R(\theta)$ is the rotation around the $z$-axis. To be more specific, we have
\beq
S(\alpha(p,\Lambda),\beta(p,\Lambda)) = e^{-i\alpha(p,\Lambda) (J^2 - K^1)}  e^{-i\beta(p,\Lambda)(-J^1-K^2)}, \qquad R(\theta(p,\Lambda))=e^{-i\theta(p,\Lambda) J^3}.
\eeq
To see this, let's take spin-1 as an example. We can set the momentum to the standard momentum $k^\mu = k (1,0,0,1)$ and take $\Lambda^\mu_{\ \nu}$ as $ S^\mu_{\ \nu} $ or $R(\theta)$, and the conditions become respectively:
\beq
\epsilon^\mu (\mathbf k, \sigma) e^{-i\sigma \theta} = R^\mu_{\ \nu} \epsilon^\nu(\mathbf k,\sigma) ,\qquad \epsilon^\mu(\mathbf k, \sigma) = S^\mu_{\ \nu} \epsilon^\nu(\mathbf k, \sigma).
\eeq
The solutions to the first equation read
\beq
\epsilon^{\mu}(\mathbf k, \pm 1 ) = \frac{1}{\sqrt{2}} (0,1,\pm i,0),
\eeq
but then the second equation can't be satisfied for general parameters $\alpha,\beta$. Instead,  applying little group transformation $S(\alpha,\beta)$ will  give us the  polarization functions as follows:
\beq
\epsilon^\mu(\mathbf k,\pm) \rightarrow \epsilon^\mu(\mathbf k,\pm) - \frac{\alpha\pm i \beta}{\sqrt{2} \, k}  k^\mu. 
\eeq
This is the origin of the necessity of gauge invariance. Nevertheless, the polarization vectors at the general momentum can still be obtained by the standard Lorentz transformation:
\beq
\epsilon^\mu(\mathbf p,\pm) = R(\hat{\mathbf p}) B(|\mathbf p|/k) \epsilon^\mu(\mathbf k, \pm)  = R(\hat{\mathbf p}) \epsilon^\mu(\mathbf k, \pm) ,
\eeq
where we have used the fact the boost along the $z$-axis doesn't affect the $x,y$ components. The resulting polarization vectors are the same as Eq.~(\ref{eq:pvml}). Under general Lorentz transformation, they transform as a vector plus an additional term proportional to the momentum:
\beq
\epsilon^\mu(\mathbf p,\pm)  \rightarrow e^{-i\theta(p,\Lambda)} \epsilon^\mu(\mathbf p,\pm)- \frac{\alpha(p,\Lambda)\pm i \beta(p,\Lambda)}{\sqrt{2} \, k}  p^\mu.
\eeq

\subsection{3-point on-shell massive amplitudes}
\begin{table}[th]
\small
\begin{center}
\begin{tabular}{ |c|c|c| } 
 \hline 
 Vertices & On-shell amplitude& high-energy limit \\ 
 \hline
$- C_{abc} \partial_{\nu} W_{\mu}^a W^{b \mu} W^{c \nu}$  & $\begin{array}{ccc} \sqrt{2} i C_{a_1 a_2 a_3}\big( \frac{ \< \bm{1}  \bm{2}\> \<  \bm{2}  \bm{3} \> [  \bm{3} \bm{1}]}{m_1m_3} \\
+\text{c.p.t.}\big)\end{array}$  & $\begin{array}{ccc}  (1^{+1}2^{-1}3^{+1}):\quad \sqrt{2} i C_{a_1 a_2 a_3} \frac{ [13]^3}{[12][23]}\\ \\
(1^{+1}2^03^0): i \sqrt{2}C_{a_1 a_2 a_3 }\frac{ (m_1^2 - m_2^2 - m_3^2) }{ 2m_2 m_3} \frac{[12][13]}{[23]}\end{array}$ \\ 
\hline
  $ -\bar{\psi}_R  \slashed{W}_a R^{a}  \psi_R -\bar{\psi}_L  \slashed{W}_a L^{a}  \psi_L$ &
  $\begin{array}{ccc}\frac{\sqrt{2}}{m_1} \left( R^{a_1}  [ \bm{1} \bm{2} ]\< \bm{1} \bm{3} \>  \right.\\
  \left.+ L^{a_1} \< \bm{1} \bm{2} \> [ \bm{1} \bm{3}] \right)\end{array}$ & $\begin{array}{ccc} 
  (1^{+1}2^{+\frac12}3^{-\frac12}):\quad R^{a_1} \frac{[12]^2}{[23]}\\ \\
    (1^{-1}2^{-\frac12}3^{+\frac12}):\quad L^{a_1}\frac{\<12\>^2}{\<23\>}\\ \\
(1^02^{+\frac12}3^{+\frac12}):\quad   -\frac{m_2 L^{a_1} - m_3 R^{a_1}}{m_1} [23] 
  \end{array}$ \\
\hline 
$F_{abi} W_{a \mu} W^{b \mu} \phi_i$ &  $2 F_{a_1 a_2 i_3} \frac{[\bm{1} \bm{2}] \< \bm{2} \bm{1} \>}{m_1 m_2}$ & $(1^{+1}2^{0}3^{0}):\quad -\sqrt{2} \frac{ F_{a_1 a_2 i_3}}{m_2} \frac{[12][13]}{[23]} $\\
\hline  
$-G_{aij} W_{a \mu} \partial^\mu  \phi_i \phi_j $ &$ \frac{ i }{\sqrt{2} m_1}  G_{a_1 i_2 i_3 } \< \bm{1} |p_2 - p_3 | \bm{1} ]$  & $ (1^{+1}2^{0}3^{0}):\quad - i \sqrt{2} G_{a_1  i_2 i_3  } \frac{[12][13]}{[23]} $\\
  \hline
  $-\frac{1}{6} P_{ijk} \phi_i \phi_j \phi_k$ & $-P_{i_1 i_2 i_3}$ & $(1^{0}2^{0}3^{0}):\quad -P_{i_1 i_2 i_3}$\\
\hline
  $  - \left( \bar{\psi}_L H_{i} \psi_R + \bar{\psi}_R H^{ \dagger}_i \psi_L \right) \phi_i$ & $ H_{i_1} [\bm{2} \bm{3}] + H^{ \dagger}_{i_1} \<\bm{2} \bm{3} \>$ &  $\begin{array}{cccc}  (1^{0}2^{+\frac12}3^{+\frac12}):\quad H_{i_1} [23] \\
  (1^{0}2^{-\frac12}3^{-\frac12}):\quad H^{ \dagger}_{i_1} \<23 \>  
   \end{array}$\\  
\hline
\end{tabular}
\end{center}
\caption{The vertices in the Lagrangian and the corresponding on-shell massive amplitudes. Here $\text{c.p.t.}$ means cyclic permutation terms.  All the momenta are taking to be ingoing. The on-shell amplitudes are entirely fixed by tree unitarity; the vertices are listed only to match the conventional normalization of the coupling constants. For on-shell massive amplitudes involving fermions, our convention is $(1_{W/\phi},2_\psi, 3_{\bar \psi})$, and we suppress the fermion internal quantum number $\feri_2$ and $\feri_3$. The last column shows the corresponding helicity amplitudes in the high energy limit.}
\label{tab:vertices}
\end{table}

We start by listing in Table~\ref{tab:vertices} all the interactions and the corresponding 3-pt tree-unitary on-shell amplitudes for an arbitrary, finite number of massive scalars $\phi_i$, fermions $\psi_\feri$ and vectors $W^a_\mu$, and present the derivations in the following.\footnote{Notice that we are using the sans serif $\{ \feri, \ferj , \cdots\}$ to denote the fermionic internal quantum numbers, to differentiate from the scalar state labels $\{ i, j, \cdots \}$.} We are adopting a purely on-shell approach in this paper, thus it is sufficient to impose tree unitarity on the complete basis of 3-pt massive amplitudes given by Ref. \cite{Durieux:2020gip}. On the other hand, we would like to make connections to the results computed using Feynman rules, thus we also calculate the same amplitudes using the Lagrangian in Ref. \cite{Cornwall:1974km} and the polarization functions derived in Section \ref{sec:polfun}, just to match the normalization of the coupling constants.

Let's start from the $WWW$ on-shell amplitudes and assume that all the vector bosons are massive. A complete independent basis including 7 different terms has been derived in Ref.~\cite{Durieux:2019eor} and the requirement of tree-level unitarity for the 3-pt on-shell amplitude $\mM_3 \lesssim \mO (E)$ has singled out the following unique structure:
\beq
\label{eq:3wos} 
 \frac{\<  \bm{1}  \bm{2}\>  \< \bm{2}  \bm{3}\> [ \bm{3}  \bm{1} ] }{m_3m_1} +  \frac{\<  \bm{2}  \bm{3}\>  \< \bm{3}  \bm{1}\> [ \bm{1}  \bm{2} ] }{m_1m_2}  +  \frac{\<  \bm{3}  \bm{1}\>  \< \bm{1}  \bm{2}\> [ \bm{2}  \bm{3} ] }{m_2m_3} ,
 \eeq
or simply
\beq
 \frac{ \< \bm{1}  \bm{2}\> \<  \bm{2}  \bm{3} \> [  \bm{3} \bm{1}]}{m_3 m_1} +\text{c.p.t.}
\eeq
where $\text{c.p.t.}$ means cyclic permutation terms. The above is clearly totally anti-symmetric in exchanging the external particle labels. Therefore, the uniqueness of the on-shell amplitude and its permutation symmetry tells us 
that  after adding the vector indices, the coupling constant $C_{a_1 a_2 a_3}$ will be completely anti-symmetric.

 This can also be seen by plugging the polarization vectors obtained in the previous section into the interaction in the Lagrangian:
\beq
- C_{abc} \partial_{\nu} W_{\mu}^a W^{b \mu} W^{c \nu},
\eeq
and the resulting amplitude reads\footnote{Our convention for the amplitudes is the same as Peskin and Schroeder~\cite{Peskin:1995ev}:
\beq
S_{\beta,\alpha} = \delta_{\beta,\alpha} + i (2\pi)^4 \delta^{(4)}(p_\alpha - p_\beta)\mM_{\beta,\alpha}.
\eeq
but with all the momenta ingoing.}
\beq
\label{eq:3WFR}
\mM_3(1^{a_1},2^{a_2},3^{a_3}) = \frac{ i}{\sqrt{2}}  \left(C_{a_1 a_2 a_3}  \frac{\< \bm{3} | p_1 | \bm{3} ] [\bm{1} \bm{2}] \< \bm{2} \bm{1} \>}{m_1 m_2 m_3} + \text{p.t.} \right).
\eeq
Here $\text{p.t.}$ means permutation terms. We first realize that the amplitude vanishes for  the symmetric part of indices $(a_1,a_2)$, i.e.:
\beq
C_{\{a_1a_2\} a_3} = 0. \label{eq:id1}
\eeq
Secondly, by using Schouten identity
\beq
| \bm{3} ] [\bm{1} \bm{2}] + | \bm{1} ] [\bm{2} \bm{3} ] + | \bm{2} ] [\bm{3} \bm{1}]=0
\eeq
and Dirac equations
\beq
p|\bm{p} ] = m | \bm{p}\>, \qquad \< \bm{p}| p = -m [ \bm{p}|,
\eeq
we can bring Eq.~(\ref{eq:3WFR}) in the form of Eq.~(\ref{eq:3wos}), and the requirement of proportionality to Eq.~(\ref{eq:3wos}) leads to
\beq
\label{eq:id2}
C_{[ab] c} = C_{[bc] a} = C_{[ca] b}.
\eeq
Combining this with Eq.~(\ref{eq:id1}) again tells us  that $C_{a_1 a_2 a_3}$ is fully anti-symmetric.  This leads to the following normalization of the on-shell $WWW$ massive amplitude:
\beq
\label{eq:3ptvvv}
\mM_3(1^{a_1}, 2^{a_2},3^{a_3})  = \sqrt{2} i C_{a_1 a_2 a_3}\left( \frac{ \< \bm{1}  \bm{2}\> \<  \bm{2}  \bm{3} \> [  \bm{3} \bm{1}]}{m_1m_3} +\text{c.p.t.}\right).
\eeq
The similar consideration of the other marginal operator that one may write down in the Lagrangian,
\beq
 -A_{abc} \ \vep_{\mu\nu\rho\sigma}  \partial^\mu W^\nu_a W^\rho_b W^\sigma_c, \qquad A_{abc}  = - A_{acb}
 \eeq
enforces the following relations:
\beq
A_{a_1 a_2 a_3} = A_{a_2 a_3 a_1} = A_{a_3 a_1 a_2},
\eeq
and as a result, the on-shell amplitude vanishes. Actually, one can verify that in this case, the Lagrangian is a total derivative. We arrive at Eq.~(\ref{eq:3ptvvv}) as our only three-vector-boson on shell massive amplitude. We will further impose that the coupling constant $C_{abc}$ to be real, as required by the  optical theorem, which demands the imaginary part of the forward  scattering amplitude  to be proportional to  the cross section to every possible final state: 
\beq
\label{eq:opth}
\text{Im}\mM_{\alpha,\alpha} \sim \sum_{\beta} \sigma(\alpha \rightarrow \beta ).
\eeq
Since the cross section usually starts at  the 3-pt coupling to the fourth order for $2 \rightarrow 2$ scattering, this relation immediately tells that the imaginary part of the of four particle amplitudes should start at loop level. By studying all possible scattering processes, it is possible to show that all coupling constants  in the 3-particle amplitudes should be the case to make the Lagrangian real. In the following discussion,  we impose these constraints on all the couplings.

We can take the high energy limit by specifying the spin components along the three-momentum direction
and the resulting polarization amplitudes are functions of $(\lambda,\tilde \lambda, \eta, \tilde \eta)$ in Eq.~(\ref{eq:msexp}). Let's take the helicity configuration  ($1^+,2^-,3^+$)  as an example:
\beq
\mM_3(1^+,2^-,3^+) = \mM_3(1^{\frac12, \frac12}, 2^{-\frac12, -\frac12}, 3^{\frac12, \frac12}) = \sqrt{2} i C_{a_1 a_2 a_3}\left( \frac{ \< \eta_1  2\> \<  2  \eta_3 \> [  31]}{m_1m_3}\right)  +\mO(m).
\eeq
By using the fact that for the total-plus 3-pt on-shell  massless amplitudes, the three angular spinors are proportional to each other, we have
\beq
\frac{[23]}{[31]}  = \frac{\langle 1\eta_1\rangle}{\langle 2 \eta_1\rangle }, \qquad \frac{[31]}{[12]}  = \frac{\langle 2\eta_3\rangle}{\langle 3 \eta_3\rangle} ,
\eeq
and it is easy to see that the helicity amplitude becomes
\beq
 \mM_3(1^+,2^-,3^+)  = \sqrt{2} i C_{a_1 a_2 a_3} \frac{ [13]^3}{[12][23]}.
\eeq
Alternatively, we can start from the above on-shell massless amplitude and invert the procedure to IR deform it to the massive case (see Appendix~\ref{app:IRdef} for detail and see also Ref. \cite{Balkin:2021dko}). In addition to the above three massless vector amplitude, the $\SU (2)$ covariant massive amplitude  in Eq.~(\ref{eq:3ptvvv}) naturally consists of massless vector-scalar-scalar amplitude. After an  involved but straightforward calculation, we can show that: 
\beq
 \mM_3(1^+,2^0,3^0)   = i \sqrt{2}C_{a_1 a_2 a_3 }\frac{ (m_1^2 - m_2^2 - m_3^2) }{ 2m_2 m_3} \frac{[12][13]}{[23]}.
\eeq

Next, we consider the $W \psi \bar{\psi}$ amplitude. The complete basis for such an amplitude is given by the following 4 terms \cite{Durieux:2020gip}:
\bea
 \< \bm{12} \> \< \bm{13} \>  ,\quad [ \bm{12} ][ \bm{13} ], \quad \< \bm{12}\> [ \bm{13} ] , \quad \< \bm{13}\> [ \bm{12} ]  ,
\eea
and it is known that the former two scale as $\mO (E^2)$ in the high energy limit \cite{Durieux:2019eor}, thus should be dropped when imposing tree unitarity; the latter two, on the other hand, scale as $\mO (E)$ and should remain. Now we can match the basis to the following interaction terms in the Lagrangian:
\beq
-\bar{\psi}_R  \slashed{W}_a R^a  \psi_R -\bar{\psi}_L  \slashed{W}_a L^{a}  \psi_L
\eeq
where $L^{a}, R^{a}$ are the Hermitian matrices in the space of fermion internal quantum numbers with labels $\{ \feri, \ferj, \cdots \}$, which we usually suppress, i.e. $\bar{\psi} L^a \psi \equiv \bar{\psi}_\feri L^a_{\feri \ferj} \psi_ \ferj$, etc. After  the substitution of the following polarization functions:
\beq
u_R(p)  =  ( 0, |\bm{p}])^T,\quad u_L(p) \to (|\bm{p} \>, 0 )^T,\quad \bar{v}_L (p) \to ( 0, [\bm{p}|), \quad \bar{v}_R (p) \to (\< \bm{p} |,0 )
\eeq
as shown in Table \ref{tab:wavefunction}, we obtain the on-shell massive amplitude as
\beq
\label{eq:vff}
\mM_3(1^{a_1}, 2_\psi, 3_{\bar \psi}) \equiv \mM_3(1^{a_1}, 2, \bar{3}) =\frac{\sqrt{2}}{m_1} \left( R^{ a_1 }  [ \bm{1} \bm{2} ]\< \bm{1} \bm{3} \> + L^{ a_1 } \< \bm{1} \bm{2} \> [ \bm{1} \bm{3}] \right).
\eeq
Note that the two terms are related by parity transformation:\footnote{ Similar to the massless spinor-helicity variables, one can  check this explicitly by using the formulae  in Eq.~(\ref{eq:lambdaeta}) and perform the parity transformation: $\theta\rightarrow \pi - \theta, \phi \rightarrow \phi + \pi$. We can  also show that under this transformation, $p_{\alpha\dot\alpha} = \lambda_\alpha^I \tilde{\lambda}_{I\dot\alpha}$ changes to $p^{\dot\alpha \alpha} =  \lambda^{I\alpha} \tilde{\lambda}_{I}^{\dot\alpha}$ which is consistent with the parity transformation on the momentum $(p^0, \vec p) \rightarrow (p^0, -\vec p)$. The change of sign of the little group index $I$ is also consistent with interpretation that it corresponds to the spin eigenstates along the momentum direction under our choice of standard Lorentz transformation.}
\beq
\lambda^I_\alpha \rightarrow i \tilde{\lambda}^{-I \dot \alpha}, \qquad  \tilde{\lambda}^{I \dot \alpha}\rightarrow i \lambda^{-I}_\alpha.
\eeq
Alternatively, one can obtain the same amplitudes by starting from the UV massless amplitudes:
\beq
\mM_{3,R}(1^{+1}, 2^{+\frac 12}, 3^{-\frac 12}) = \frac{[12]^2}{[23]}, \qquad \mM_{3,L} (1^{-1}, 2^{-\frac 12}, 3^{+\frac 12}) = \frac{\<12\>^2}{\<23\>}
\eeq
and  following the procedure outlined in Appendix~\ref{app:IRdef} to IR deform them to the massive on-shell amplitudes. It also can be shown that the IR-unified on-shell massive amplitude in Eq.~(\ref{eq:vff}) contains the following UV fermion-fermion-scalar massless amplitude:
\beq
\mM_3 (1^0,2^{+\frac 12},3^{+\frac12}) = -\frac{m_2 L^{ a_1 } - m_3 R^{ a_1 }}{m_1} [23] .
\eeq
Note that the coupling factor is proportional to $m_\psi/m_W$, which is indeed in the form that one would expect from the Higgs mechanism. This is consistent with the understanding that Higgs mechanism can be thought as IR-unification of different massless UV amplitudes~\cite{Arkani-Hamed:2017jhn}.

Now we turn to the interaction terms involved scalars. The $WW\phi$ amplitude has the following 3-term basis:
\bea
 \< \bm{12} \>^2,\quad [ \bm{12} ]^2 ,\quad \< \bm{12}\> [ \bm{21} ]  ,
\eea
where only the last term satisfies tree unitarity, and it is symmetric in exchanging the two vector labels, thus the corresponding (real) coupling constant $F_{a_1 a_2 i_3}$ needs to be symmetric in $\{a_1, a_2 \}$. The $W\phi\phi$ amplitude has a 1-term basis
\bea
\< \bm{1} | p_2 - p_3 | \bm{1} ]
\eea
which already satisfies tree unitarity; it is anti-symmetric in the two external scalar labels, thus the associated (real) coupling constant $G_{a_1 i_2 i_3}$ needs to be anti-symmetric in $\{i_2, i_3 \}$. The $\phi \phi \phi$ amplitude has to be a (real) constant $P_{i_1 i_2 i_3}$, which satisfies tree unitarity and needs to be totally symmetric.

On the other hand, the $\phi \psi \bar{\psi}$ amplitude has the following basis
\bea
\< \bm{23} \>  , \quad  [ \bm{23} ],
\eea
where both terms satisfy tree unitarity. We can write down the following amplitude:
\bea
\mM_3(1^{i_1}, 2, \bar{3})=H_{i_1} [\bm{2} \bm{3}] + H^\dagger_{i_1} \<\bm{2} \bm{3} \>,
\eea
where we have suppressed the indices $\{ \feri_2, \feri_3 \}$ for the fermionic internal quantum numbers of $\{ \psi_{\feri_2}, \bar{ \psi}_{\feri_3} \}$, i.e. $H_{i_1} \equiv (H_{i_1})_{\feri_2 \feri_3}$ etc. The coupling constants in front of $\< \bm{23} \>$ and $ [ \bm{23} ] $ are related by  Hermitian conjugation because of the aforementioned optical theorem of Eq. (\ref{eq:opth}). The relevant three-particle operators in the Lagrangian is as follows:
\beq
 F_{abi} W_{a \mu} W^{b \mu} \phi_i -G_{aij} W_{a \mu} \partial^\mu  \phi_i \phi_j - \frac{1}{3!} P_{ijk} \phi_i \phi_j \phi_k- \left( \bar{\psi}_L H_{i} \psi_R + \bar{\psi}_R H^{ \dagger}_i \psi_L \right) \phi_i.
\eeq
It is straightforward to derive the on-shell massive amplitudes from the above, which fixed the normalization of the coupling constants as given by Table~\ref{tab:vertices}.

\section{Four particle amplitudes and the tree-level unitarity}
\label{sec:4pt}
Now we construct 4-pt amplitudes from unitarity and locality.  Locality tells us that when one internal momentum is going on-shell, the amplitudes have simple poles  in terms of Mandelstam variables, and unitarity  requires that the residue is  the product of lower-pt amplitudes. To be more explicit, one can write the 4-pt amplitudes as
\bea
\label{eq:amp4}
\mM_4 = \mM_{4,\txtf} + \mM_{4,\txtc},
\eea
where $\mM_{4,\txtf}$ contains the non-local parts of different factorization channels,  while $\mM_{4,c}$ are the possible additional contact terms. The latter is a linear combination of all local terms given the particle contents, expressed in the stripped-contact-term (SCT) basis $\{\mM_{4,\txtc}^{(i)} \}$ given by Ref. \cite{Durieux:2020gip}:
\bea
\mM_{4,\txtc} = \sum_{i} c_i \mM_{4,\txtc}^{(i)},
\eea
where $c_i$ are polynomials of Mandelstam variables. Apparently, the slowest energy growing behavior for these terms are achieved when $c_i$ are constants. On the other hand, the factorizable part $\mM_{4,\txtf}$ is fixed by unitarity:
\bea
\label{eq:ampfact}
 \mM_{4,\txtf} = - \sum_{\mI}\sum_{i=2}^4 \frac{1}{s_{1i}^2-m_\mI^2}  \mM_{3,i_L}^{\{I_1, I_2, \cdots , I_{2s_\mI} \}}  \ep_{I_1 J_1} \cdots \ep_{I_{2s_\mI} J_{2s_\mI}}  \mM_{3,i_R}^{\{J_1, J_2, \cdots , J_{2s_\mI} \}},
\eea
 where $s_{ij} \equiv (p_i + p_j)^2$, and we sum over all possible states $\mI$ of mass $m_\mI$ and spin $s_\mI$ as well as all possible factorization channels.  Here again, we take all the momenta as ingoing, which means that in the real momentum limit, some of the momenta have negative energy. We have the following analytical continuation:
  \beq
  \label{eq:neg}
 \lambda^I(-p) =- \lambda^I(p), \qquad \tilde{\lambda}_I (-p) =  \tilde{\lambda}_I (p).
\eeq
In the above convention, the 3-point amplitudes $\mM_{3,i_{L}}$ has momenta $ p_1$, $p_i$ and $-p_1 - p_i$, while $\mM_{3,i_R}$ has momenta $\{p_j \}$ with $j \in \{1,2,3,4\} \setminus \{1, i\}$ and $p_1 + p_i$. Notice that in Sec.~\ref{sec:3pt},  on-shell massive amplitudes are considered equivalent if they are related by equations of, but different forms of 3-pt on-shell amplitudes certainly lead to different formulae for the local terms with different coefficients $c_i$.

In order to obtain the coefficient $c_i$ and the coupling relations, we will take the high energy limit of the amplitude $\mM_4$ at fixed, non-zero angles and  impose the tree-level unitary criterion, which requires that the energy growing behavior of the four-particle amplitude should be at most a constant. 
As discussed in detail in Appendix~\ref{app:msv}, we can expand the massive spinors for the external states in the little group space as
\bea
\lambda_\alpha^I = \lambda_\alpha \zeta^{-I} + \eta_\alpha \zeta^{+I},\qquad \lat^I_{\dota} = \lat_\dota \zeta^{+I} + \ett_\dota \zeta^{-I},
\label{eq:etadef}
\eea
and the helicity amplitudes for particle with spin $S$ in a particular frame can be obtained by extracting the coefficients of  $\left((\zeta^+)^{S+h}(\zeta^-)^{S-h}\right)^{I_1\cdots I_{2s}}$. The resulting helicity amplitudes are functions of $(\lambda_i, \tilde \lambda_i, \eta_i,\tilde \eta_i)$ with explicit formulae as follows:
\bea
&&\lambda_{i,\alpha} = \sqrt{E_i + p_i} \left( \begin{array}{c}
-s_i^*\\
c_i
\end{array} \right), \qquad \lat_{i,\dota} = \sqrt{E_i + p_i} \left( \begin{array}{c}
-s_i \\
c_i
\end{array} \right),\nn\\
&&\eta_{i,\alpha} = \sqrt{E_i - p_i} \left( \begin{array}{c}
c_i^*\\
s_i
\end{array} \right), \qquad \ett_{i,\dota} =- \sqrt{E_i - p_i} \left( \begin{array}{c}
c_i \\
s_i^*
\end{array} \right),
\eea
where $E_i$ and $p_i$ are the energy and the magnitude of 3-momentum for each external particle $i$, and $c_i, s_i$ are defined as (see the Appendix~\ref{app:mshv} for the discussion of the phase convention)
\bea
c_i = \cos \frac{\theta_i}{2} e^{\frac i 2\phi_i}, \qquad s_i = \sin \frac{\theta_i}{2} e^{\frac i 2\phi_i}. 
\eea
We have assumed that the energy of the particle is positive and for negative energy, the spinors are obtained by the analytic continuation in Eq.~(\ref{eq:neg}) and in all cases, $E_i =+ \sqrt{ m_i^2 + p_i^2}$. To simplify the derivation, we will work in the center-of-mass frame, which is obtained by setting the angles as follows:
\bea
&\theta_1 = 0,\qquad \theta_2 = \pi, &\qquad \theta_3  = \theta, \qquad \theta_4 = \pi - \theta,\nn\\
&\phi_1 = \phi_2 = 0, &\qquad \phi_3 = \phi, \qquad \phi_4 = \phi + \pi,
\eea
The magnitude of the 3-momenta $p_i$ can be obtained using momentum conservation and on-shell condition as
\bea
&&p_1 = p_2 = E \sqrt{\left(1-\frac{m_1^2 + m_2^2}{4E^2}\right)^2  - \frac{m_1^2 m_2^2}{E^4}},\nn\\
&&p_3 = p_4 = E \sqrt{\left(1-\frac{m_3^2 + m_4^2}{4E^2}\right)^2  - \frac{m_3^2 m_4^2}{E^4}},
\eea
where $2 E = \sqrt{s_{12}}$ is the total center-of-mass energy. All the helicity amplitudes will be functions of energy $E$ and scattering angles $(\theta, \phi)$. We will take the fixed-angle high energy limit  and extract the tree-level unitary conditions by setting  to zero the coefficients of linearly independent functions of $(\theta, \phi)$ with energy growing behavior faster than $\mO(E^0)$. Table 1 in Ref. \cite{Durieux:2020gip} shows the SCT basis for the contact terms, organized by helicity components in which these contact terms have the fastest energy growing behavior. We will see that only the contact terms corresponding to dimension-4 operators in an EFT  Lagrangian will survive the tree unitarity constraints. 

\subsection{Constraining the 4-pt amplitudes}

Let us enumerate all the possible 4-pt amplitudes.

\subsubsection{$WWWW$}

We start with the $WWWW$ amplitude. As discussed above, the factorizable part of the amplitude $\mM_{4,\txtf}$ can be obtained by gluing the 3-pt on-shell massive amplitudes together and adding back the simple pole structures $1/({s_{ij} - m_\mI^2})$. By naive energy scaling counting, the energy growing behavior of $\mM_{4,\txtf}$ is  at most $\mO (E^4)$, while  in contrast the contact terms of helicity components \amp+000, \amp+++0, and \amp++-0 are at least $\mO (E^5)$, thus the coefficients $c_i$ of these contact terms must vanish. One example of such contact terms is $[\1 \2] [\3 \4] \< \2 \4 \1 ] \< \3 \4 \>$, which is in the \amp+000 category. The $\mO(E^4)$ energy growing behavior of $\mM_{4,\txtf}$ arises  only from \amp0000 helicity category. This eliminates the possibility of adding contact terms for helicity configurations \amp++00, \amp+-00, \amp++++ and \amp+++-, as they are at least $\mO (E^4)$. On the other hand, we need \amp0000 contact terms in $\mM_{4,\txtc}$, which can be  parameterized as
\bea
 &&[\1 \2][\3 \4] \left(c_{W^4,1} \<\1 \2\> \<\3 \4\> + c_{W^4,2} \<\1 \3\> \<\2 \4\>  \right) \nn\\
 &&+ [\1 \3][\2 \4] \left(c_{W^4,3} \<\1 \2\> \<\3 \4\> + c_{W^4,4} \<\1 \3\> \<\2 \4\>  \right).
\eea
The requirement that $\mM_4 = \mM_{4,\txtf} + \mM_{4,\txtc}$ satisfies tree unitarity then uniquely fix the coefficients $c_{W^4,i}$. This means that we have completely determined the form of $\mM_{4,\txtc}$. (The only helicity configuration for the contact terms that we have not discussed is \amp+++-, where contact terms in the category simply cannot exist.) The amplitude is determined to be
\bea
&&\mM_4 (1^{a_1},2^{a_2},3^{a_3},4^{a_4}) \nn\\
&=& \mM_{4,s} (1^{a_1},2^{a_2},3^{a_3},4^{a_4}) + \mM_{4,s} (2^{a_2},3^{a_3},1^{a_1},4^{a_4}) + \mM_{4,s} (3^{a_3},1^{a_1},2^{a_2},4^{a_4}),\label{eq:amp4v}
\eea
where the $s$-channel component is
\bea
\mM_{4,s} (1^{a_1},2^{a_2},3^{a_3},4^{a_4}) &=& - \sum_{b }\frac{C_{a_1 a_2 b} C_{a_3 a_4 b} N_{W^4,b} }{ m_{a_1} m_{a_2} m_{a_3} m_{a_4} (s_{12} - m_{b}^2 )} \nn\\
&&- \sum_{i } \frac{4F_{a_1 a_2 i} F_{a_3 a_4 i } \< \1 \2 \> \<\3 \4\> [\1 \2]  [\3 \4]}{m_{a_1} m_{a_2} m_{a_3} m_{a_4} (s_{12} - m_{i}^2 )},
\eea
and
\bea
N_{W^4,b} &=&  \left(\<\3\4\3]  \<\4 (\2 - \1) \4]  + \<\4\3\4]   \<\3(\1 - \2) \3] \right)  \<\1 \2\> [\1 \2]\nn\\
&&+ \left( \<\2\1\2]   \<\1 (\3 - \4) \1]  + \<\1\2\1]  \<\2 (\4 - \3) \2]  \right)  \<\3 \4\> [\3 \4]\nn\\
&&+2 \left( \<\2\1\2] (\<\3\4\3] \<\1 \4\> [\1 \4] - \<\4\3\4] \<\1 \3\> [\1 \3])  + \<\1\2\1] (\<\4\3\4] \<\2 \3\> [\2 \3] - \<\3\4\3] \<\2 \4\> [\2 \4])\right)\nn\\
&&+\< \1 \2 \> \<\3 \4\> [\1 \2]  [\3 \4] \left( \frac{(m_{a_1}^2 - m_{a_2}^2)( m_{a_3}^2 - m_{a_4}^2)}{m_{b}^2} + s_{12} + 2 s_{13} \right.\nn\\
&&\left.\phantom{\frac{(m_{a_1}^2)}{m_{b}^2}}- m_{a_1}^2 - m_{a_2}^2 - m_{a_3}^2- m_{a_4}^2\right)\nn\\
&&+\left( \< \1 \3 \> \<\2 \4\> [\1 \3]  [\2 \4] - \< \1 \4 \> \<\2 \3\> [\1 \4]  [\2 \3]\right) (s_{12} - m_{b}^2 ).
\eea
The $t$- and $u$-channels are given by $\mM_{4,s} (2^{a_2},3^{a_3},1^{a_1},4^{a_4})$ and $ \mM_{4,s} (3^{a_3},1^{a_1},2^{a_2},4^{a_4})$, respectively. The terms on the RHS of Eq. (\ref{eq:amp4v}), as well as similar expressions below, are organized according to the internal states of the factorization channels, as indicated by the particle label that is summed, or the masses in the propagators. For example, the first term on the RHS of  Eq. (\ref{eq:amp4v}) has a sum of vector index $b$, and together with the mass $m_b^2$ indicate that this is the contribution of a vector boson exchange. Notice that we have absorbed the contact terms into the different factorization channels in a symmetric way.

Next, the $\mO (E^3)$ terms need to vanish as well. Here it is convenient to calculate in the center of mass frame. We find that $\mO (E^3)$ terms only exist for \amp+000. This leads to the following constraint on the coefficient $C_{abc}$ of the $WWW$ amplitudes:
\bea
C_{ a_1 a_2 b} C_{a_3 a_4 b} + C_{ a_1 a_3 b} C_{a_4 a_2 b} + C_{ a_1 a_4 b} C_{a_2 a_3 b} = 0,\label{eq:jacobi}
\eea
which has the form of the Jacobi identity. Therefore, we can identify the totally anti-symmetric $C_{abc}$ as the structure constants for some  compact Lie group $G$.

We then proceed to consider the $\mO (E^2)$ terms, which only exist for \amp0000. This leads to the following constraint for $C_{abc}$ as well as the coefficients of $WW\phi $ amplitudes $F_{abi}$:
\bea
&&-4 \left(F_{a_1 a_3 i } F_{a_2 a_4 i } -  F_{a_1 a_4 i } F_{a_2 a_3  i } \right) \nn\\
&=& \sum_{b} \left\{ \left( C_{ a_1 a_3 b} C_{a_2 a_4 b}  - C_{ a_1 a_4 b} C_{a_2 a_3 b} \right)  (3 m_{b}^2 - m_{a_1}^2 - m_{a_2}^2 - m_{a_3}^2 - m_{a_4}^2)\right.\nn\\
&& + \frac{1}{m_{b}^2} \left[C_{ a_1 a_3 b} C_{a_2 a_4 b} (m_{a_1}^2 - m_{a_3}^2)( m_{a_2}^2 - m_{a_4}^2)  \right.\nn\\
&&\left.\left.-C_{ a_1 a_4 b} C_{a_2 a_3 b} (m_{a_1}^2 - m_{a_4}^2)( m_{a_2}^2 - m_{a_3}^2) \right] \right\}.\ \ \label{eq:vvvvrel}
\eea
This relation agrees with Eq. (7) in Ref.~\cite{Cornwall:1973tb}. One can check explicitly that as long as the constraints in Eqs. (\ref{eq:jacobi}) and (\ref{eq:vvvvrel}) are satisfied, the amplitude in Eq. (\ref{eq:amp4v}) behaves as $\mO (E^0)$: we have extracted all information given by tree unitarity.

\subsubsection{$WWW\phi$}
We next consider the $WWW\phi$ amplitude.  In this case, the factorizable part $\mM_{4,\txtf}$ at high energy limit grows at most at $\mO (E^2)$, but  the energy growing behavior of the contact terms $\mM_{4,\txtc}$, if non-vanishing, are  at least of $\mO (E^3)$. Therefore, we don't need the contact terms in this case, and the amplitude is fully determined by the factorizable part $\mM_{4,\txtf}$:
\bea
&&\mM_4 (1^{a_1},2^{a_2},3^{a_3},4^{i}) \nn\\
&=& \mM_{4,s} (1^{a_1},2^{a_2},3^{a_3},4^{i}) + \mM_{4,s} (2^{a_2},3^{a_3},1^{a_1},4^{i}) + \mM_{4,s} (3^{a_3},1^{a_1},2^{a_2},4^{i})\label{eq:ampv3s},
\eea
where the $s$-channel component is given by
\bea
&&\mM_{4,s} (1^{a_1},2^{a_2},3^{a_3},4^{i})\nn\\
 &=& - i\sqrt{2} \left(\sum_{b }\frac{C_{a_1 a_2 b} F_{b a_3 i} N_{W^3\phi,b} }{ m_{a_1} m_{a_2} m_{a_3}  (s_{12} - m_{b}^2 )}+\sum_{j }   \frac{ 2F_{a_1 a_2 j} G_{a_3 ji } \<\3\4\3] \<\1 \2\> [\1 \2] }{m_{a_1} m_{a_2} m_{a_3} ( s_{12} - m_{j}^2 )}  \right),
\eea
with
\bea
N_{W^3\phi,b} &=&  \<\3(\1 - \2)\3] \<\1 \2\> [\1 \2]+2\left( \<\1\2\1] \<\2 \3\> [\2 \3]- \<\2\1\2] \<\1 \3\> [\1 \3] \right)\nn\\
&&+ \frac{m_{a_1}^2- m_{a_2}^2}{m_b^2} \<\3\4\3] \<\1 \2\> [\1 \2].
\eea

Now, tree unitarity requires that the $\mO (E^2)$ terms in Eq. (\ref{eq:ampv3s}) vanish, which turn out to only exist for the \amp0000 component. This leads to the following constraints on $C_{abc}$, $F_{abi}$ as well as the coefficient for the $W\phi\phi$ amplitude $G_{aij}$:
\bea
&&\sum_b \frac{1}{2 m_{b}^2} \left[ C_{a_2 a_3 b} (m_{b}^2 + m_{a_2}^2- m_{a_3}^2) F_{b a_1 i } - C_{ a_2 a_1 b} (m_{b}^2 + m_{a_2}^2 -m_{a_1}^2 ) F_{a_I a_3 i } \right]\nn\\
&&   =   F_{a_2 a_3 j } G_{a_1   i j } - F_{a_1 a_2 j } G_{a_3   i j } - C_{a_1 a_3 b}  F_{b a_2 i } .\label{eq:vvvsrel}
\eea
One can check that upon this constraint, the amplitude in Eq. (\ref{eq:ampv3s}) is $\mO (E^0)$. 

\subsubsection{$WW\phi\phi$}
We then turn to the  $WW \phi \phi$ amplitude.  In the high energy limit, the factorizable part $\mM_{4,\txtf}$ is of $\mO (E^2)$, while contact terms for helicity components \amp+000 and \amp+-00 start at $\mO (E^3)$ and $\mO(E^4)$, respectively, thus they are eliminated by tree unitarity. The contact terms for \amp++00 start at $\mO (E^2)$, but terms in $\mM_{4,\txtf}$ that are in this helicity category are only $\mO (E^0)$, thus contact terms for \amp++00 cannot exist either. On the other hand, for the \amp0000 component $\mM_{4,\txtf}$ is $\mO (E^2)$, while the contact terms of $\mO (E^2)$  can be parameterized as
\bea
&&c_{W^2 \phi^2} \<\1 \2\> [\1 \2] .
\eea
The requirement that the $\mO (E^2)$ contributions of the full amplitude vanish completely determines the coefficient $c_{W^2 \phi^2}$, so that the total amplitude is calculated to be
\bea
&&\mM_4 (1^{a_1},2^{a_2},3^{i_3},4^{i_4})\nn\\
&=& \sum_{b } \left( \frac{C_{a_1 a_2 b} G_{b i_3 i_4 } N_{W^2\phi^2,b} }{ m_{a_1} m_{a_2}   (s_{12} - m_{b}^2 )} + \frac{ 2F_{b a_1 i_3} F_{b a_2 i_4} (\< \1 \3 \1] \< \2 \4 \2] - 2m_b^2  \<\1 \2 \> [\1 \2] )}{m_{a_1} m_{a_2} m_{b}^2 ( s_{13} - m_{b}^2 )} \right.\nn\\
&&\left. + \frac{ 2F_{b a_1 i_4} F_{b a_2 i_3} (\< \1 \4 \1] \< \2 \3 \2] -  (s_{23} + m_b^2)  \<\1 \2 \> [\1 \2] )}{m_{a_1} m_{a_2} m_{b}^2 ( s_{23} - m_{b}^2 )} \right)+\sum_{j }   \left(-\frac{ 2F_{a_1 a_2 j} P_{ ji_3 i_4 } \<\1 \2\> [\1 \2] }{m_{a_1} m_{a_2}  ( s_{12} - m_{j}^2 )}  \right. \nn\\
&& \left. + \frac{ 2G_{ a_1j i_3} G_{ a_2 j i_4}  \< \1 \3 \1] \< \2 \4 \2]  }{m_{a_1} m_{a_2}   ( s_{13} - m_{j}^2 )} + \frac{ 2G_{ a_1j i_4} G_{ a_2 j i_3} (\< \1 \4 \1] \< \2 \3 \2] -  (s_{23} - m_j^2)  \<\1 \2 \> [\1 \2] )}{m_{a_1} m_{a_2}  ( s_{23} - m_{j}^2 )}\right),\label{eq:ampv2s2}
\eea
with
\bea
N_{W^2\phi^2,b} &=&  \left( \frac{(m_{a_1}^2 - m_{a_2}^2)(m_{i_3}^2 - m_{i_4}^2)}{m_b^2} - m_b^2+ m_{a_1}^2 + m_{a_2}^2+ m_{i_3}^2 + m_{i_4}^2 - 2s_{23} \right) \<\1 \2 \> [\1 \2] \nn\\
&&+ 2(\< \1 \3 \1] \< \2 \1 \2] - \< \1 \2 \1] \< \2 \3 \2] ).
\eea
However, fixing $c_{W^2 \phi^2}$ is necessary but not sufficient to make the $\mO (E^2)$ terms vanish in the above; we need an additional constraint on  the coupling constants $C_{abc}$, $F_{abi}$ and $G_{aij}$:
\bea
 -\sum_b \frac{1}{ m_{b}^2} \left(  F_{ a_1 b i_3 } F_{ a_2 b i_4 } -  F_{b a_1 i_4 } F_{b a_2 i_3 } \right)  =  G_{a_1   i_3 j } G_{a_2   i_4 j } - G_{a_1  i_4 j } G_{a_2   i_3 j } +C_{ a_1 a_2 b} G_{b  i_3 i_4}  , \label{eq:vvssrel}
\eea
which agrees with Eq. (8) in Ref. \cite{Cornwall:1973tb}. One can check that the constraint above will make the amplitude in Eq. (\ref{eq:ampv2s2}) satisfy tree unitarity. Notice that no constraint has been put on the coefficient $P_{ijk}$ for the $\phi^3$ amplitude. Actually, in order to obtain non-trivial constraint on the pure-scalar interactions, one need go to higher-pt amplitudes~\cite{Chang:2019vez,Falkowski:2019tft}.

\subsubsection{$W\psi W \bar{\psi}$}
We now turn to the 4-pt amplitudes involving fermions. First, we consider the case of $W\psi W \bar{\psi}$. In the high energy limit at fixed, non-zero angle, the factorizable part $\mM_{4,\txtf}$ is growing at most at $\mO (E^2)$ while a non-vanishing $\mM_{4,\txtc}$ would be at least $\mO (E^3)$. Therefore, the possible contact terms are forbidden by tree unitarity, and the amplitude is fully determined by the factorizable part:
\bea
&&\mM_4 (1^{a_1},2^{\feri_2},3^{a_3},\bar{4}^{\feri_4})\nn\\
&=& \sum_{b }  \frac{iC_{a_1 a_3 b}  N_{W^2\psi^2,b} }{ m_{a_1} m_{a_3}   (s_{13} - m_{b}^2 )} + \sum_i \frac{2F_{a_1 a_3 i} \<\1 \3 \> [\1 \3 ] \left( (H_i)_{\feri_4 \feri_2}[\2\4] + (H_{i}^{\dagger})_{\feri_4 \feri_2} \< \2 \4 \> \right)}{ m_{a_1} m_{a_3}   (s_{13} - m_{j}^2 )}\nn\\
&&-\frac{2}{m_{a_1} m_{a_3}} \sum_{\ferj }   \left(\frac{1}{  ( s_{23} - m_{\ferj}^2 )} \left[ L^{a_1}_{\feri_4 \ferj} L^{a_3}_{\ferj \feri_2} \<\2 \3\> [\1 \4] ( [\1 \3] m_{a_1} - \< \1 \4 \3] )\right.\right.\nn\\
&&\left.+ R^{a_1}_{\feri_4 \ferj} R^{a_3}_{\ferj \feri_2} \<\1 \4\> [\2 \3] ( [\1 \3] m_{a_3} + \< \3 \2 \1] ) + m_{\ferj}  \left( L^{a_1}_{\feri_4 \ferj}  R^{a_3}_{\ferj \feri_2}  \<\1 \3\>[\1\4 ] [\2 \3] +R^{a_1}_{\feri_4 \ferj}  L^{a_3}_{\ferj \feri_2}  \<\1 \4\> \<\2 \3\> [\1 \3] \right) \right]\nn\\
&&\left. \phantom{\frac{1}{m_{\ferj}^2}} + 1\leftrightarrow 3 \right),\label{eq:ampv2f2}
\eea
with the numerator factor as follows:
\bea
N_{W^2\psi^2,b} &=& L^b_{\feri_4 \feri_2} \left( \< \2 (\1 - \3) \4]\<\1 \3 \> [\1 \3] + 2 \left( \< \3 \1 \3]\<\1 \2 \> [\1 \4] + \< \1 \3 \1]\<\2 \3 \> [\3 \4]\right) \phantom{\frac{m_{a_3}^2 }{m_b^2}}\right.  \nn\\
&&\left.+ \frac{m_{a_1}^2-m_{a_3}^2 }{m_b^2} \<\1 \3 \> [\1 \3] \left( m_{\feri_4}\<\2 \4 \> - m_{\feri_2} [\2 \4] \right)\right) \nn\\
&&+R^b_{\feri_4 \feri_2} \left( \< \4 (\1 - \3) \2]\<\1 \3 \> [\1 \3] - 2 \left( \< \3 \1 \3]\<\1 \4 \> [\1 \2] + \< \1 \3 \1]\<\3 \4 \> [\2 \3]\right) \phantom{\frac{m_{a_3}^2 }{m_b^2}}\right.  \nn\\
&&\left.+ \frac{m_{a_1}^2-m_{a_3}^2 }{m_b^2} \<\1 \3 \> [\1 \3] \left( m_{\feri_4} [\2 \4 ] - m_{\feri_2} \<\2 \4\> \right)\right) .
\eea
The $\mO (E^2)$ terms only exist for \amp0-0+ and \amp0+0-, and for them to vanish we arrive at the following constraints:
\bea
iC_{a_1 a_3 b} L^b = [L^{a_1}, L^{a_3}], \qquad iC_{a_1 a_3 b} R^b = [R^{a_1}, R^{a_3}].\label{eq:lrla}
\eea
As  $C_{abc}$ has been identified in Eq. (\ref{eq:jacobi}) as structure constants in some Lie group $G$, the above commutation relations indicates that $L^a$ and $R^a$ are generators in some representations of $G$.

Upon identifying the commutation relations, the amplitude in Eq. (\ref{eq:ampv2f2}) behaves as $\mO (E^0)$ for all helicity components except for \amp0+0+, which are $\mO (E)$. Tree unitarity then imposes another constraint
\bea
&& 2 F_{ a_1  a_3 i } (H_i)_{\feri_4 \feri_2} - m_{\feri_2} \left\{L^{a_1}, L^{a_3} \right\}_{\feri_4 \feri_2}  - m_{\feri_4} \left\{ R^{a_1}, R^{a_3} \right\}_{\feri_4 \feri_2}+\sum_{\ferj} 2  m_{\ferj} \left( L^{a_1}_{\feri_4 \ferj} R^{a_3}_{\ferj \feri_2} + L^{a_3}_{\feri_4 \ferj} R^{a_1}_{\ferj \feri_2} \right)  \nn\\
&=&  \sum_b i C^{a_1 a_3 b} \frac{\left(m_{a_1}^2 - m_{a_3}^2 \right)}{m_{b}^2} \left(m_{\feri_2} L^{b}_{\feri_4 \feri_2} - m_{\feri_4} R^{b}_{\feri_4 \feri_2}    \right) ,\label{eq:vvffrel}
\eea
which ensures that the full amplitude is $\mO (E^0)$.

\subsubsection{$W\psi \phi \bar{\psi}$}
Next, we study  the $W\psi \phi \bar{\psi}$ amplitude. Similar to the above case, the full amplitude is fully determined by the factorizable part $\mM_{4,\txtf}$, which 
 in the high energy limit is  growing at most at $\mO (E)$. On the other hand, $\mM_{4,\txtc}$ is at least $\mO (E^2)$, thus set to $0$ by tree unitarity. The amplitude is listed as follows:
\bea
&&\mM_4 (1^{a},2^{\feri_2},3^{i},\bar{4}^{\feri_4})\nn\\
&=&     \sum_{\ferj} \frac{\sqrt{2}}{m_a} \left( \frac{  1 }{ s_{12} - m_\ferj^2 } \left[ \<\1\2\> \left( m_\ferj [\1 \4](H_i)_{\feri_4 \ferj} + (m_{\feri_4} [\1 \4] + \< \4 \3\1 ]) (H_i^{\dagger})_{\feri_4 \ferj}\right) L^a_{\ferj \feri_2} \right. \right.\nn\\
&&\left.+ [\1\2] \left( m_\ferj \<\1 \4 \>(H_i^{\dagger})_{\feri_4 \ferj} + (m_{a_1} [\1 \4] - \< \1 \2\4 ]  ) (H_i)_{\feri_4 \ferj}\right) R^a_{\ferj \feri_2} \right]\nn\\
&&+\frac{  1 }{ s_{23} - m_\ferj^2 } \left[ L^a_{\feri_4 \ferj} [\1\4] \left( m_\ferj \<\1 \2 \> (H_i^{\dagger})_{\ferj \feri_2} + (m_{a_1} [\1 \2] - \< \1 \4\2 ]) (H_i)_{\feri_4 \ferj}\right)  \right. \nn\\
&&\left. \phantom{\frac{  1 }{   m_\ferj^2 }}\left.+ R^a_{\feri_4 \ferj}\< \1\4 \> \left( m_\ferj [\1 \2 ] (H_i)_{\ferj \feri_2} + (m_{\feri_4} [\1 \2] + \< \2 \3\1 ]  ) (H_i^{\dagger})_{\ferj \feri_2} \right)  \right] \right)\nn\\
&&+ \sum_b \frac{\sqrt{2} F_{bai} }{ m_{a} m_b^2    (s_{13} - m_{b}^2 )} \left[ L^b_{\feri_4 \feri_2} \left(2 m_b^2 \< \1 \2 \> [ \1 \4 ] + \< \1 \3 \1 ] ( m_{\feri_2} [\2 \4] - m_{\feri_4} \< \2 \4 \>) \right) \right.\nn\\
&& \left. + R^b_{\feri_4 \feri_2} \left(2 m_b^2 \< \1 \4 \> [ \1 \2]+ \< \1 \3 \1 ] ( m_{\feri_2} \<\2 \4 \> - m_{\feri_4} [ \2 \4 ]) \right) \right]\nn\\
&& +\sum_j \frac{ \sqrt{2} iG_{aji} \<\1 \3\1] \left( (H_j)_{\feri_4 \feri_2}[\2\4] +(H_{j}^{\dagger})_{\feri_4 \feri_2} \< \2 \4 \> \right)}{ m_{a}    (s_{13} - m_{j}^2 )}.\label{eq:vsffa}
\eea
The $\mO (E)$ contributions come from the \amp0+0+ helicity components, and for them to vanish we need to impose the following relation:
\bea
 \sum_b \frac{1}{m_b^2} F_{ a b  i   } ( m_{\feri_4} R^{b}_{\feri_4 \feri_2} -m_{\feri_2} L^{b}_{\feri_4 \feri_2}  ) =    iG_{a   i j }  (H_j)_{\feri_4 \feri_2} - ( L^{a} H_i )_{\feri_4 \feri_2} +  (H_i R^{a} )_{\feri_4 \feri_2} ,\label{eq:vsffrel}
\eea
which ensures that the full amplitude is tree unitary.

\subsubsection{Amplitudes for other processes}

For all of the other 4-particle processes, $\mM_{4,\txtf}$ is already $\mO (E^0)$, thus the only possible contact term is the constant term in the $\phi \phi \phi \phi$ amplitude. There are no non-trivial relations obtained in these processes, but for completeness, we list them below. First,   the  amplitude for $W \phi \phi \phi$ is given by
\bea
&&\mM_4 (1^{a},2^{i_2},3^{i_3},4^{i_4}) \nn\\
&=& \mM_{4,s}(1^{a},2^{i_2},3^{i_3},4^{i_4}) + \mM_{4,s} (1^{a},4^{i_4},2^{i_2},3^{i_3}) + \mM_{4,s} (1^{a},3^{i_3},4^{i_4},2^{i_2}) ,
\eea
with s-channel contribution as
\bea
\mM_{4,s} (1^{a},2^{i_2},3^{i_3},4^{i_4}) &=& i\sqrt{2} \left( \sum_b \frac{F_{bai_2} G_{bi_3 i_4} \left( 2 \< \1 \3 \1] m_b^2 + \< \1 \2 \1 ] (m_b^2 + m_{i_3^2} - m_{i_4}^2) \right)}{m_b^2 m_a (s_{12} - m_b^2)} \right. \nn\\
&&\left. - \sum_j \frac{ G_{a j i_2} P_{ji_3i_4} \< \1 \2 \1 ] }{m_a (s_{12} - m_j^2)} \right).
\eea

The $\phi \phi \phi \phi$ amplitude reads
\bea
&&\mM_4 (1^{i_1},2^{i_2},3^{i_3},4^{i_4}) \nn\\
&=& \mM_{4,s} (1^{i_1},2^{i_2},3^{i_3},4^{i_4}) + \mM_{4,s} (2^{i_2},3^{i_3},1^{i_1},4^{i_4}) + \mM_{4,s} (3^{i_3},1^{i_1},2^{i_2},4^{i_4}) - K_{i_1 i_2 i_3 i_4},\quad 
\eea
where the s-channel contribution is given by:
\bea
&&M_{4,s} (1^{i_1},2^{i_2},3^{i_3},4^{i_4})\nn\\
& = & -\sum_a \frac{G_{a i_1 i_2} G_{a i_3 i_4} \left[ m_a^2 (s_{13} + m_a^2 - m_{i_1}^2 - m_{i_2}^2 - m_{i_3}^2 - m_{i_4}^2) + ( m_{i_1}^2- m_{i_2}^2)(m_{i_3}^2- m_{i_4}^2) \right]  }{m_b^2 (s_{12} - m_a^2)}\nn\\
&&-\sum_j \frac{ P_{i_1 i_2 j} P_{i_3 i_4 j}}{(s_{12} - m_j^2)},
\eea
and $K_{i_1 i_2 i_3 i_4}$ is the constant contact term, which needs to be totally symmetric because of Bose symmetry.

Finally, we  have the $\phi \psi \phi \bar{\psi}$ amplitude as
\bea
&&M_4 (1^{i_1},2^{\feri_2},3^{i_3},\bar{4}^{\feri_4})\nn\\
&=& \sum_\ferj \left[ \frac{1}{s_{12} - m_\ferj^2} \left( (H_{i_3})_{\feri_4 \ferj}  (H_{i_1}^\dagger)_{\ferj \feri_2} ( m_{\feri_2} [\2 \4] - \< \2 \1 \4 ] ) + (H_{i_3}^\dagger)_{\feri_4 \ferj}  (H_{i_1})_{\ferj \feri_2} ( m_{\feri_4} [\2 \4] + \< \4 \3 \2 ] )\right. \right.\nn\\
&&\left. \phantom{\frac{1}{ m_\ferj^2}} \left. (H_{i_3} )_{\feri_4 \ferj}  (H_{i_1})_{\ferj \feri_2}   m_{\feri_2} [\2 \4] + (H_{i_3}^\dagger)_{\feri_4 \ferj}  (H_{i_1}^\dagger)_{\ferj \feri_2} ( m_{\feri_2} \< \2 \4 \>\right) + 1 \leftrightarrow 3 \right]\nn\\
&&- \sum_a \frac{iG_{ai_1 i_3}}{m_a^2 (s_{13} - m_a^2)} \left[ L^a_{\feri_4 \feri_2} \left(2 m_a^2 \<  \2  \1 \4 ] -(m_a^2 + m_{i_1}^2 - m_{i_3}^2) ( m_{\feri_2} [\2 \4] - m_{\feri_4} \< \2 \4 \>) \right) \right.\nn\\
&& \left. + R^a_{\feri_4 \feri_2} \left(2 m_a^2 \<  \4  \1 \2] -(m_a^2 + m_{i_1}^2 - m_{i_3}^2) ( m_{\feri_2} \<\2 \4 \> - m_{\feri_4} [ \2 \4 ]) \right) \right]\nn\\
&&+ \sum_j \frac{ P_{ji_1 i_3} \left( (H_{j})_{\feri_4 \feri_2}  [\2 \4 ] + (H_{j}^\dagger)_{\feri_4 \feri_2}  \< \2 \4 \>\right)}{(s_{13} - m_j^2)},
\eea
and the $\psi  \bar{\psi}\psi  \bar{\psi}$ amplitude reads:
\bea
&&M_4 (1^{\feri_1},\bar{2}^{\feri_2},3^{\feri_3},\bar{4}^{\feri_4})\nn\\
&=& \left( \sum_a \frac{1}{ m_a^2 ( s_{12} - m_a^2) } \left[ \left(L^a_{\feri_4 \feri_3}  ( m_{\feri_3} [ \3 \4 ] - m_{\feri_4} \< \3 \4 \> ) +R^a_{\feri_4 \feri_3} ( m_{\feri_3} \< \3 \4 \> - m_{\feri_4} [ \3 \4 ] )  \right)   \right. \phantom{\frac{\left(] (H_i^\dagger)_{\feri_2 } \right) }{] m_i^2}}\right.\nn\\
&&\times \left( L^a_{\feri_2 \feri_1} ( m_{\feri_1} [ \1 \2 ] - m_{\feri_2} \< \1 \2 \> ) + R^a_{\feri_2  \feri_1}  ( m_{\feri_1} \< \1 \2 \> - m_{\feri_2} [ \1 \2 ] ) \right) \nn\\
&&\left.- 2 m_a^2 \left( L^a_{\feri_4 \feri_3} R^a_{\feri_2 \feri_1}   \<  \2 \3 \> [ \1 \4 ] + L^a_{\feri_2 \feri_1} R^a_{\feri_4 \feri_3}     \<  \1 \4 \> [ \2 \3 ]  +  L^a_{\feri_2 \feri_1} L^a_{\feri_4 \feri_3}  \<  \1 \3 \> [ \2 \4 ]   +  R^a_{\feri_2 \feri_1} R^a_{\feri_4 \feri_3}  \<  \2 \4 \> [ \1 \3 ] \right) \right] \nn\\
&& \left.- \sum_i \frac{\left( (H_i)_{\feri_2 \feri_1} [ \1 \2 ] + (H_i^\dagger)_{\feri_2 \feri_1} \< \1 \2 \> \right) \left( (H_i)_{\feri_4 \feri_3} [ \3 \4 ] + (H_i^\dagger)_{\feri_4 \feri_3} \< \3 \4 \> \right) }{s_{12} - m_i^2} \right) + 1 \leftrightarrow 3.
\eea

\subsection{Interpretation of the constraints}
\label{sec:interp}
We see that tree unitarity completely fixes the 4-pt amplitudes in terms of 3-pt amplitudes, with the exception of the additional parameter $K_{ijkl}$ as the constant scalar contact term. Moreover, tree unitarity puts additional constraints to all parameters of 3-pt amplitudes except for   the $\phi \phi \phi$ interaction $P_{ijk}$. We obtain the relations in Eqs. (\ref{eq:jacobi}), (\ref{eq:vvvvrel}), (\ref{eq:vvvsrel}), (\ref{eq:vvssrel}), (\ref{eq:lrla}), (\ref{eq:vvffrel}) and (\ref{eq:vsffrel}). It is easy to see that Eqs. (\ref{eq:jacobi}) and (\ref{eq:lrla}) indicate that the totally antisymmetric coupling constants $C_{abc}$ are structure constants of some Lie group $G$, and the $W\psi  \bar{\psi}$ coupling matrix $L^a$ and $R^a$ are generators of some representations of the same  Lie group. The other relations put constraints on $WW\phi$ couplings $F_{abi}$, $W\phi \phi$ couplings $G_{aij}$ and the $\phi \psi  \bar{\psi}$ couplings $H_i$. To see the meaning of these relations clearly, we define the following coupling matrices:
\bea
T^a_{ij} = iG_{aij} ,\qquad T^a_{ib} = -T^a_{bi} = \frac{i}{m_b}  F_{abi}, \qquad T^a_{bc} = i C_{abc} \frac{m_a^2 - m_b^2 - m_c^2 }{2m_b m_c}.\label{eq:brbos}
\eea
Then Eqs. (\ref{eq:vvvvrel}), (\ref{eq:vvvsrel}), (\ref{eq:vvssrel}) become
\bea
iC_{abe} T^e_{cd} &=& T^a_{c \tilde{k}} T^b_{\tilde{k} d} - T^b_{c \tilde{k}} T^a_{\tilde{k} d},\\
i C_{abd} T^d_{ic} &=& T^a_{i \tilde{k}} T^b_{\tilde{k} c} - T^b_{i \tilde{k}} T^a_{\tilde{k} c},\\
iC_{abc} T^c_{ij} &=& T^a_{i \tilde{k}} T^b_{\tilde{k} j} - T^b_{i \tilde{k}} T^a_{\tilde{k} j},
\eea
where the index $\tilde{k}$ runs over both the vector indices $\{a \}$ and the scalar indices $\{ i \}$. We will see that this corresponds to all of the scalar states in the UV, including the longitudinal components of the massive vector bosons. This motivates us to group $T^a_{ij}$, $T^a_{ib}$ and $T^a_{bc}$ together as an anti-symmetric matrix $T^a_{\tilde{i} \tilde{j}}$, and the above is just
\bea
i C_{abc} T^c_{\tilde{i} \tilde{j}} = [T^a, T^b]_{\tilde{i} \tilde{j}}.\label{eq:unicom}
\eea
In other words, the interactions between the vector and scalar states together form a  generator of the representation of the Lie group $G$. If a generator $a$ belongs to the Abelian invariant subgroup of $G$, the structure constant $C_{abc}$ vanishes for all $b,c$ and we can have additional St\ddu ckelberg mass terms for the corresponding Abelian vector bosons.

Similarly, for the fermion Yukawa interactions, one can generalize $H_i$ to $H_{\tilde{i}}$, by the following definition when $\tilde{i} $ is a vector index $a$:
\bea
 \left( H_a \right)_{\feri \ferj} \equiv \frac{i}{m_a} \left(  m_\ferj L^a  - m_{\feri} R^a  \right)_{\feri \ferj} .\label{eq:brfer0}
\eea
After the extension, the relations in
Eqs. (\ref{eq:vvffrel}) and (\ref{eq:vsffrel}) then become
\bea
L^a H_{b} - H_{b} R^a -  H_{i} T^a_{ib} - H_{c} T^a_{cb} &=& 0,\label{eq:fr12}\\
L^a H_i -  H_i R^a -H_j T^a_{ji}  -  H_{b}  T^a_{bi} &=& 0,\label{eq:fr11}
\eea
which can be combined into the following identity:
\bea
L^a H_{\tilde{i}} - H_{\tilde{i}} R^a - H_{\tilde{j}} T^a_{\tilde{j} \tilde{i}} = 0.\label{eq:brfer1}
\eea
This tells us that the coupling matrices $(H_{\tilde{i}})_{\feri \ferj} $ are rank-3 invariant tensors of Lie group $G$, where the indices $\tilde{i}$, $\feri$ and $\ferj$ transform in the representation associated with $T^a$, $L^a$ and $R^a$, respectively. It means that the 3-particle on-shell amplitudes involving the fermions, physical scalar states and the longitudinal components of the massive gauge boson respect the symmetry generated by Lie group $G$.

One can understand the definitions in Eqs. (\ref{eq:brbos}) and (\ref{eq:brfer0}) at the Lagrangian level, which we discuss in Appendix \ref{app:basrel}; see also Refs. \cite{Brod:2019bro,Bishara:2021buy} where similar relations are derived using current conservation. However, there is a much more straightforward, on-shell way to arrive at these definitions. In Table \ref{tab:vertices} we presented the 3-pt UV massless amplitudes contained in the high energy limit of the IR massive amplitudes. In particular, the same massless vector-scalar-scalar helicity amplitude in the UV can be generated by 3 different massive amplitudes in the IR:
\bea
\label{eq:vssml}
 \mM_3(1^+,2^0,3^0)   = \left\{ \begin{array}{ll}
i \sqrt{2}C_{a_1 a_2 a_3 }\frac{ (m_1^2 - m_2^2 - m_3^2) }{ 2m_2 m_3} \frac{[12][13]}{[23]} & \text{from } \mM_3(1^{a_1}, 2^{a_2},3^{a_3})\\
- \sqrt{2} \frac{ F_{a_1 a_2 i_3}}{m_2} \frac{[12][13]}{[23]} & \text{from } \mM_3(1^{a_1}, 2^{a_2},3^{i_3})\\
- i \sqrt{2} G_{a_1  i_2 i_3  } \frac{[12][13]}{[23]} & \text{from } \mM_3(1^{a_1}, 2^{i_2},3^{i_3})
\end{array}\right. ,
\eea
i.e. the longitudinal modes of the vector states in the IR amplitudes can be identified with scalar external states in the UV. Now, in the spirit of the Goldstone boson equivalence theorem~\cite{Cornwall:1974km,Lee:1977eg,Chanowitz:1985hj}, we want to unify all the scalar states in the UV, including the physical scalar and the longitudinal components of vector states in the IR, under a universal coupling and a single group representation of the gauge symmetry, then the redefinition in Eq. (\ref{eq:brbos}) is completely natural. Note that the definition of Eq. (\ref{eq:brbos}) has also taken into account the factor of $-i$ between the amplitude of longitudinal component of massive vector boson in the high energy limit and corresponding Goldstone boson amplitude. Conversely, our ability to use the redefinition in Eq. (\ref{eq:brbos}) to arrive at the unified commutation relation of Eq. (\ref{eq:unicom}) suggests a spontaneously broken symmetry, as the (longitudinal components of) vector states and scalar states are clearly distinct in the IR, which are only unified in the UV. Similarly, for 3-pt amplitudes involving fermions, the massless scalar-fermion-fermion amplitude in the UV can be obtained by two distinct massive amplitudes in the IR:
\bea
\mM(1^0,2^{+\frac 12},3^{+\frac12}) =\left\{ \begin{array}{ll}
- \frac{m_2 L^{ a_1 } - m_3 R^{ a_1 }}{m_1} [23] & \text{from } \mM_3(1^{a_1}, 2_\psi,3_{\bar \psi})\\
H_{i_1} [23] & \text{from } \mM_3(1^{i_1}, 2_\psi,3_{\bar \psi})
\end{array}  \right. .
\eea
Again, imposing the Goldstone boson equivalence in the UV makes the definition in Eq. (\ref{eq:brfer0}) completely natural, which again manifests the existence of a spontaneously broken symmetry that unifies the vector and scalar states.

One can also compare our general setting with the special case of the electroweak theory in the SM. For example, Ref. \cite{Bachu:2019ehv} studied the 4-pt bosonic amplitudes in the electroweak theory. As we are considering all external states to be massive, to compare with their results we need to decouple the photons in Ref. \cite{Bachu:2019ehv}, i.e. setting the coupling to photons $e = 0$, and as a result the $W$ and $Z$ boson have the same mass: $m_Z = m_W$. Then relevant coupling constants are identified as
\bea
-\frac{1}{\sqrt{2}} C_{abc}  \to  e_W \text{ for $W^+ W^- Z$}, \quad
-\frac{2F_{abi}}{m_a m_b}  \to  \left\{ \begin{array}{ll}
\frac{e_{ZZH}}{m_Z} & \text{for $ZZh$}\\
\frac{e_{WWH}}{m_W} & \text{for $W^+ W^- h$}
\end{array}, \right.\quad
G_{aij}  \to  0,
\eea
and the tree-unitarity relations in  Eq. (\ref{eq:vvvvrel}) and Eq. (\ref{eq:vvvsrel})  becomes
\bea
e_{WWH}^2 = 2 e_W^2, \qquad e_{WWH} e_{ZZH} = 2 e_W^2, \qquad e_{WWH} = e_{ZZH}.
\eea
Our results of the bosonic 4-pt amplitudes as well as the above relations agree with Ref.~\cite{Bachu:2019ehv}. We see that the above relations are the result of a spontaneously broken symmetry, where the vector and scalar states are unified under the same group representation.

Another example is Ref. \cite{Durieux:2019eor}, which, in addition to the electroweak sector in SM, also considered a single generation of fermions, and studied the 4-pt $\psi^c \psi Z h$ amplitude, which is a special case of our consideration with the  following  values for the coupling constants:
\bea
\frac{2}{m_a} F_{abi} \to -c^{00}_{ZZh},&\dfrac{\sqrt{2}}{m_a} R^a \to \dfrac{1}{m_Z} c^{LR0}_{\psi^c \psi Z},&\frac{\sqrt{2}}{m_a} L^a \to \frac{1}{m_Z} c^{RL0}_{\psi^c \psi Z},\non\\
G_{aij} \to 0, &H_i \to - c^{RR}_{\psi^c \psi h},& H^{\dagger}_i \to - c^{LL}_{\psi^c \psi h}.
\eea
Then Eq. (\ref{eq:vsffrel}) becomes
\bea
\left( c^{LR0}_{\psi^c \psi Z} - c^{RL0}_{\psi^c \psi Z} \right) \left(\frac{m_\psi}{ 2 m_Z } c^{00}_{ZZh} - c^{RR}_{\psi^c \psi h} \right) = 0.
\eea
Again, our amplitude in Eq. (\ref{eq:vsffa}) agrees with Ref. \cite{Durieux:2019eor} upon the proper identification of the coupling constants, and we agree on the above relation as well. We see clearly that the above relation comes from the constraint that the Yukawa coupling needs to be an invariant tensor, again a consequence of a spontaneously broken symmetry. 

\section{Conclusion and outlook}
\label{sec:conclusion}
\begin{table}[thb]
\centering
\begin{tabular}{|c|c|c|c|}
\hline
Particles& Couplings&  Processes &  Relations\\
\hline
 \multirow{2}{*}{  $WWW$} &\multirow{2}{*}{$C_{abc}$} & $W^{(\pm)}W^{(0)}W^{(0)}W^{(0)}~\mO(E^3)$& Jacobi identity, Eq.~(\ref{eq:jacobi}) 
\\
\cline{3-4}
 &  & $\begin{array}{cc}W^{(0)}W^{(0)}W^{(0)}W^{(0)}~\mO(E^2)\\W^{(0)}W^{(0)}W^{(0)}\phi~~\mO(E^2) \\
W^{(0)}W^{(0)}\phi\phi~~\mO(E^2) \end{array}$ & $\begin{array}{c}
  C_{abc} \sim \frac{m_b m_c}{m_a^2 - m_b^2 - m_c^2 } T^a_{bc}, \text{ Lie algebra for } T^a, \\
\text{Eqs. (\ref{eq:vvvvrel}), (\ref{eq:vvvsrel}) and (\ref{eq:vvssrel}) }\end{array}$\\
\hline
$WW\phi$ & $F_{abi}$ & $\begin{array}{cc}W^{(0)}W^{(0)}W^{(0)}W^{(0)}~\mO(E^2)\\W^{(0)}W^{(0)}W^{(0)}\phi~~\mO(E^2) \\
W^{(0)}W^{(0)}\phi\phi~~\mO(E^2) \end{array}$ & $\begin{array}{c}
F_{abi} \sim m_a T^{b}_{ia}, \text{ Lie algebra for } T^a, \\
\text{Eqs. (\ref{eq:vvvvrel}), (\ref{eq:vvvsrel}) and (\ref{eq:vvssrel}) }\end{array}$\\
\hline
$W\phi\phi$ & $G_{aij}$ & $\begin{array}{cc}W^{(0)}W^{(0)}W^{(0)}\phi~~\mO(E^2) \\
W^{(0)}W^{(0)}\phi\phi~~\mO(E^2) \end{array}$& $\begin{array}{c}
G_{aij} \sim  T^a_{ij}, \text{ Lie algebra for } T^a\\
\text{Eqs. (\ref{eq:vvvsrel}) and (\ref{eq:vvssrel}) }
\end{array}$ \\
\hline
$W \psi \bar \psi$ &$L^a_{\feri \ferj}, R^a_{\feri \ferj}$&  $W^{(0)}\psi^{(\pm)} W^{(0)}\bar{\psi}^{(\mp)}~\mO(E^2)$ &   Lie algebra for $L^a, R^a$, Eq.~(\ref{eq:lrla}) \\
\hline
$\phi \psi \bar \psi $ & $H^i_{\feri\ferj}$ &  $\begin{array}{ccc}W^{(0)}\psi^{(\pm)}W^{(0)}\bar{\psi}^{(\pm)}~\mO(E)\\
W^{(0)} \psi^{(\pm)} \phi \bar{\psi}^{(\pm)}~\mO(E)\end{array}$ &  $\begin{array}{c}
H^i_{\feri \ferj} \text{ part of an invariant tensor,}\\
 \text{Eqs.~(\ref{eq:vvffrel}) and (\ref{eq:vsffrel})}
\end{array} $\\
\hline
\end{tabular}
\caption{Summary of couplings, processes and the corresponding relations considered in the paper.  The superscripts in the particle type labels in the processes indicate the helicities of the corresponding particles in the high energy limit, and we also indicate the energy growing behaviors for each case. Relations among the coupling constants and the masses are schematically displayed in the last column.  }
\label{tab:summary}
\end{table}
In this paper, we have considered the most general 3-pt on-shell massive amplitudes with energy scaling at most $\mO(E)$, involving  an arbitrary, finite number of scalar, spinor and vector  particle states defined as irreducible representations of the little group. Starting from 3-pt on-shell amplitudes, we have calculated the full 4-point amplitudes from unitary and locality, which lead to the formulae to construct the 4-pt on-shell amplitudes in Eq.~(\ref{eq:amp4}) and Eq.~(\ref{eq:ampfact}). The contact terms are further determined by the requirement of tree unitarity, which states that the energy growing behaviors of $n$-pt amplitudes in the fixed-angle high energy limit should not exceed $\mO(E^{4-n})$. For 4-pt amplitudes, the leading energy growing behavior should be at most a constant. Moreover, the requirement of tree unitarity further imposes relations on the 3-pt couplings constants and the masses of the particles. In Table~\ref{tab:summary}, we summarize the processes and relations obtained in this approach and they coincide  with Ref.~\cite{Cornwall:1974km}.  We can see that the fastest energy growing behaviors happen in the longitudinal modes of the massive vectors, which is consistent with the fact that the St\ddu eckelberg  scalars are always associated with derivatives. As discussed in Section~\ref{sec:interp} and Appendix~\ref{app:basrel}, the relations can be understood from the point of view of the Lie algebra. This includes the Jacobi identity for the triple-vector couplings, commutation relations for vector-fermions couplings, and the predictions of the Higgs mechanism for the scalar-vector and scalar fermion couplings. They all converge to the gauge invariance from the UV interactions with possible modifications by the vector mass terms of the invariant Abelian subgroups. 

From the study, we have shown that the on-shell massive and massless amplitudes are manifestly little group covariant and they further unleash the power of  quantum mechanics and special relativity. With analytical continuation into complex momenta, we are able to discuss 3-pt on-shell massive spin amplitudes and massless helicity amplitudes. The requirement of little group covariance puts strong constraints on the possible structures and especially for the massless case, it uniquely determines the helicity amplitudes. In this paper, we have seen that the tree unitarity at the 3-pt on-shell massive amplitudes level already constrains the coupling constants, like that 3-vector couplings $C_{abc}$ should be fully antisymmetric. As also illustrated in Appendix~\ref{app:IRdef}, 3-pt on-shell  massive amplitudes can be obtained from the IR-deformation of the corresponding massless helicity amplitudes, and one interesting observation is  that the spurious poles in the massless vector amplitudes turn into vector  mass singularities in the on-shell massive amplitudes. It further induces the energy growing behaviors in the longitudinal modes of the massive vectors. In other words,  this translates the requirement of  consistent factorization for the 4-pt massless amplitudes into the requirement of no faster energy growing behaviors than it should be in the tree unitary theory.

Our study can be generalized in several ways. Firstly, one can go beyond the 4-pt scattering amplitudes and determine the form of the scalar potential from tree unitarity. As we have seen, tree unitarity at 4-pt does not impose any constraints on the $\phi \phi \phi$ coupling or the $\phi \phi \phi \phi$ contact term apart from being totally symmetric, and the relations that they satisfy can only be derived at the 5-pt level. A computation of all 5-pt processes when all external states are bosonic, and at least one of them is a scalar, should fully determine the relations satisfied by the scalar self-interactions. Secondly, it would be nice to explore how Higgsless theories can be embedded in the on-shell formalism and in that cases, no scalar degrees of freedom are involved and one needs Kaluza-Klein towers of massive vectors and fermions~\cite{Csaki:2003dt,Csaki:2003zu}. Finally, one can try to include the  massive spin-3/2 and spin-2 particles to see what non-trivial tree-unitary theory can be obtained.

Last but not least, efforts have been made to extend the color-kinematics duality and the modern double copy program  to include massive gauge bosons \cite{Chiodaroli:2015rdg,Chiodaroli:2017ehv,Chiodaroli:2018dbu,Momeni:2020vvr,Johnson:2020pny,Gonzalez:2021bes}. The variety  of coupling relations that we present here greatly extends the meaning of ``color'' relations in the usual sense of color-kinematics duality,  which may help us understand the possible double copy structures of spontaneously broken gauge theories.


\section*{Acknowledgments}
We would like to thank  Henrik Johansson, Markus Luty, Yael Shadmi, and Lian-Tao Wang for valuable  discussions. We also thank Hsin-Chia Cheng, Ian Low and John Terning for reading and commenting on the draft. The work of D.L. is supported in part by the U.S. Department of Energy under grant DE-SC-0009999. The work of Z.Y. is supported by  the Knut and Alice Wallenberg Foundation under grants KAW 2018.0116 and KAW 2018.0162.


\appendix

\section{Notations and conventions}
\label{app:notation}
In this appendix, we collect the notations and conventions used throughout the paper. We will use the mostly minus metric:
\beq
\eta_{\mu\nu} = \text{diag}(1,-1,-1,-1).
\eeq
Our momenta are parametrized as
\beq
\begin{split}
p^\mu &= (E, \vec{p}) =   (E, p_x, p_y, p_z) \\
\end{split}.
\eeq
Note that we will take all the momenta ingoing, which means that $E$ can be either positive or negative. The matrix generators of the Lorentz group in the vector representation are given as follows~\cite{Peskin:1995ev}:
\beq
(\mathcal{J}^{\mu\nu})^{\rho}_{\ \sigma} = i (\eta^{\rho\mu}\delta^\nu_\sigma - \delta^\mu_\sigma \eta^{\rho \nu})
\eeq
To be more explicit, the expressions for the the rotation generators $J^1 = \mathcal{J}^{23}, J^2 = \mathcal{J}^{31},J^3 = \mathcal{J}^{12}$ and the boost generators $K^i = \mathcal{J}^{0i}$ read:
\beq 
\label{eq:lggenerators}
\begin{split}
J^1 &= - i \left(\begin{array}{cccc}
0 & 0 & 0 &0\\
0 & 0 & 0 &0\\
0 & 0 & 0 &1\\
0 & 0 & -1 &0\\
\end{array}\right), \qquad  J^2 = - i \left(\begin{array}{cccc}
0 & 0 & 0 &0\\
0 & 0 & 0 &-1\\
0 & 0 & 0 &0\\
0 & 1 & 0 &0\\
\end{array}\right),\qquad J^3 = - i \left(\begin{array}{cccc}
0 & 0 & 0 &0\\
0 & 0 & 1 &0\\
0 & -1& 0 &0\\
0 & 0 & 0 &0\\
\end{array}\right),\\
K^1 &=  i \left(\begin{array}{cccc}
0 & 1 & 0 &0\\
1 & 0 & 0 &0\\
0 & 0 & 0 &0\\
0 & 0 & 0 &0\\
\end{array}\right), \qquad 
 K^2 =  i \left(\begin{array}{cccc}
0 & 0 & 1 &0\\
0 & 0 & 0 &0\\
1 & 0 & 0 &0\\
0 & 0 & 0 &0\\
\end{array}\right),\qquad
 K^3 = i \left(\begin{array}{cccc}
0 & 0 & 0 &1\\
0 & 0 & 0 &0\\
0 & 0& 0 &0\\
1 & 0 & 0 &0\\
\end{array}\right).\\
\end{split}
\eeq
The finite Lorentz transformation can be obtained by the exponential mapping:
\beq
\Lambda^\rho_{\ \sigma} = (e^{-\frac i2 \omega_{\mu\nu}\mathcal{J}^{\mu\nu}})^\rho_{\ \sigma}
\eeq
with the rotation angles as $\theta^1 = \omega_{23}, \cdots$ and the boost parameters (rapidities) as $\eta^i = \omega_{0i}$.
For the spinor representation in the Weyl basis of Dirac matrices, the generator matrices are:
\beq 
J^k = \frac12 \left(\begin{array}{cc}
\sigma^k& 0\\
0 &\sigma^k
\end{array}\right), \qquad K^k = -\frac i2 \left(\begin{array}{cc}
\sigma^k& 0\\
0 &-\sigma^k
\end{array}\right).
\eeq
We know that  $\SL (2,\mC)$  is the double cover of the proper, orthochronous Lorentz group $\SO (3,1)$, similar to the fact that  the $\SU (2)$ is the double cover of the rotation group $\SO (3)$~\cite{CORNWELL1997193}. This can be seen by defining the $2\times2$ matrix for each four-momentum:
\beq
p_{\alpha\dot{\alpha}} = p_\mu \sigma^\mu_{\alpha\dot{\alpha}},  \qquad p^{\dot{\alpha} \alpha} =  p_\mu \bar{\sigma}^{\mu \dot{\alpha}\alpha},
\eeq
where the sigma matrices are defined as
\beq
\sigma^\mu = (\mathbf{1}_{2\times 2}, \vec{\sigma}), \qquad  \bar \sigma^\mu = (\mathbf{1}_{2\times 2}, -\vec{\sigma}),
\eeq
with $\sigma^i$ as Pauli matrices. The four-momentum vector can be obtained by exploring the following identity~\cite{Dreiner:2008tw}:
\beq
\text{Tr}[\sigma^\mu\bar{\sigma}^\nu] = \text{Tr}[\bar{\sigma}^\mu \sigma^\nu] = 2 \eta^{\mu\nu},
\eeq
which yields
\beq
p^\mu = \frac 12 p_{\alpha\dot\alpha}\bar{\sigma}^{\mu\dot\alpha\alpha} = \frac 12 p^{\dot\alpha\alpha} \sigma^\mu_{\alpha\dot\alpha}.
\eeq
From the same identity, it can also be shown that the determinant of the momentum matrix gives the scalar product of the momentum:
\beq
\qquad \text{det} \,p_{\alpha\dot{\alpha}}  = \frac{1}{2}\varepsilon^{\alpha\beta}\varepsilon^{\dot{\alpha}\dot{\beta}} p_{\alpha \dot{\alpha}} p_{\beta\dot{\beta}}= p_\mu p^\mu ,
\eeq
which can be generalized to any two momentum vectors:
\beq
 p_{1\alpha\dot{\alpha}}   p_2^{ \dot{\alpha}\alpha}  =  2 p_1 \cdot p_2 .
\eeq
For any $\mL \in \SL (2,\mC)$, the momentum matrix  $p_{\alpha \dot\alpha}$ transforms as
\beq
\begin{split}
p &\rightarrow \mathcal{L}^\dagger \,p \, \mathcal{L}
\end{split},
\eeq
which leaves the determinant invariant. This establishes the connection between Lorentz group and $\SL (2,\mC)$. We can also see that $\mL$ and $-\mL$ gives the same Lorentz transformation. The $\SL (2,\mC)$ indices  $\alpha, \dot \alpha$ can be raised or lowered by the antisymmetric tensor $\varepsilon^{\alpha \beta}$ and its inverse $\varepsilon_{\alpha\beta}$:
\beq
\varepsilon^{12} = \varepsilon_{21} =1, \qquad  \varepsilon_{12} = \varepsilon^{21}=  -1, \qquad \varepsilon_{\alpha\beta}\varepsilon^{\beta\gamma} = \delta_\alpha^{\gamma},
\eeq
and the same definition applies to $\varepsilon^{\dot\alpha \dot \beta}, \varepsilon_{\dot\alpha \dot \beta}$.

\section{Massless spinor-helicity variables}
\label{app:mshv}
For massless particles, $p^2 = 0$, and the matrix $p_{\alpha\dot\alpha}$  has rank one, which can be always factorized as direct product of two spinors:
\beq
p_{\alpha\dot{\alpha}} =   \lambda_\alpha \tilde{\lambda}_{\dot{\alpha}}.
\eeq
For real momenta in the Minkowski space, $p_{\alpha\dot{\alpha}}$ is Hermitian, and we have
\beq
\label{eq:energy}
\tilde{\lambda}_{\dot{\alpha}} = \pm (\lambda_\alpha)^*,
\eeq
 with the sign determined by whether the energy is taken to be positive ($+$) or negative ($-$).  It is clear that the helicity variables $\lambda, \tilde{\lambda}$ satisfy the massless Weyl equations:
 \beq
 p_{\alpha\dot\alpha}\tilde \lambda^{\dot \alpha} = 0, \qquad p^{\dot \alpha \alpha} \lambda_\alpha = 0.
 \eeq
 From the definition,  it is also clear that given a particular momentum $p$, $\lambda$ and $\tilde \lambda$ are not uniquely determined but up to a scaling:
 \beq
 \lambda \rightarrow w \lambda, \qquad  \tilde \lambda \rightarrow w^{-1} \tilde \lambda ,
 \eeq
 with $w \in \mC$ being a non-zero complex number. In fact, there is no continuous way to define $\lambda$ as a function of $\vec{p}$~\cite{Witten:2003nn}, as will be seen later on from the concrete formulae. The angular and square spinor products are defined as follows:
\beq
\begin{split}
\label{eq:proddef}
  \left<12\right>  & \equiv \left<\lambda_1 \lambda_2\right> =  \lambda^{\alpha}_1\lambda_{2 \alpha} = \varepsilon_{\alpha\beta}  \lambda^{\alpha}_1\lambda^{\beta}_2,\\
[12] & \equiv [\tilde{\lambda}_1 \tilde{\lambda}_2] = \tilde{\lambda}_{1\dot{\alpha}}\tilde{\lambda}^{\dot{\alpha}}_2 = \varepsilon_{\dot{\alpha}\dot{ \beta}} \tilde{\lambda}^{1\dot{\beta}}\tilde{\lambda}^{2 \dot{\alpha}}.  \\
\end{split}
\eeq
For particle $i$, we also define the ``half-brackets''~\cite{Durieux:2019eor}:
\beq
|i\rangle = \lambda_{i\alpha}, \qquad \langle i | = \lambda_{i}^{\alpha}, \qquad |i] = \tilde{\lambda}_{i}^{\dot{\alpha}}, \qquad [i| = \tilde{\lambda}_{i \dot{\alpha}},
\eeq
and the spinor products can also be understood as follows:
\beq
  \left<12\right>  =   \left<1|^\alpha |2\right>_\alpha, \qquad [12] = [1|_{\dot\alpha}|2]^{\dot\alpha}.
\eeq

Note that in our convention, for the real momenta with same sign of energy, we have the following relation:
\beq
  \left<12\right> = - [12]^*,
\eeq
as can be directly verified by using the definition Eq.~(\ref{eq:proddef}) and Eq.~(\ref{eq:energy}). By using the fact that any fully anti-symmetric rank-2 tensor is proportional to the Levi-Civita tensor $\varepsilon$, we can obtain the identities
\beq
\lambda_{1[\alpha} \lambda_{2\beta]} = \lrag{12} \varepsilon_{\alpha\beta}, \qquad \tilde\lambda_{1[\dot\alpha} \tilde\lambda_{2\dot\beta]} =- [12] \varepsilon_{\dot\alpha\dot\beta}.
\eeq

We have some useful identities:
\beq
   \left<ij\right> [ji]  =     \left<i|p_j|i\right]  = 2 p_i \cdot p_j,
\eeq
and the Schouten-identity:
\beq
\begin{split}
&\lag 12\rag |3\rag + \lag 23\rag  |1\rag  + \lag 31\rag  |2\rag   = 0\\
&[ 12 ] |3] + [23] |1]  + [31]  |2]   = 0,\\
\end{split}
\eeq
which can be proved by using the fact that the spinor space is two-dimensional and any two spinors can provide a basis as long as their  angular/square inner product is not vanishing. 

For the parametrization $(E, \theta,\phi)$ in the real momenta (let us assume the energy is positive for the moment, $E >0$.),
\beq
 p_x = E \sin\theta \cos\phi, \qquad p_y = E \sin\theta\sin\phi, \qquad p_z = E \cos\theta.
\eeq
Then
\beq
\begin{split}
p_{\alpha\dot{\alpha}}=  \left(
\begin{array}{cc}
E (1-\cos\theta)& -E \sin\theta e^{-i\phi} \\
-E \sin\theta e^{i\phi}   &E(1+\cos\theta),
\end{array}
\right) = 2E \left(
\begin{array}{cc}
 s s^*& -c^* s^*  \\
-cs   & c c^*
\end{array}
\right) 
\end{split},
\eeq
where we have defined
\beq
 c\equiv \cos\frac\theta 2 e^{\frac i 2 \phi}, \qquad s \equiv \sin\frac\theta 2\, e^{\frac i2\phi}.
\eeq
We can choose the spinor-helicity variables as\footnote{Here we have adopted a phase convention commonly used in Quantum Mechanics~\cite{Cohen-Tannoudji:101367} and as shown below, this phase convention leads to the polarization vectors with phases $e^{\pm i\phi}$.}
\beq
\label{eq:lambda}
\lambda_\alpha = \sqrt{2 E}  \left(
\begin{array}{c}
-s^*\\
c 
\end{array}
\right), \qquad \tilde{\lambda}_{\dot{\alpha}} = \sqrt{2 E} \left(
\begin{array}{c}
 -s\\
c^*
\end{array}
\right).
\eeq
It is straightforward to verify that the spinors $\lambda_\alpha, \tilde{\lambda}^{\dot\alpha}$are the eigenvectors of the helicity operator with eigenvalues of $-\frac12 (+\frac12)$:\footnote{Note that $\dot\alpha$ is the upper index.}
\beq
h_\mO = \frac 12 \vec{\sigma} \cdot \hat{\vec{p}} = \frac12  \left(
\begin{array}{cc}
\cos\theta& \sin\theta e^{-i\phi} \\
 \sin\theta e^{i\phi}   &-\cos\theta
\end{array}
\right),
\eeq
which confirms that $\lambda(\tilde\lambda)$ carries helicity weight $-\frac12 (+\frac12)$. The presence of  functions $\sin\frac\theta 2$ and $\cos\frac \theta 2$ indicates that the spinor-helicity variables are not continuous function of the momenta.

For massless spin-1 particles, the polarization vectors can be written as
 \beq
\epsilon^-_{\alpha\dot{\alpha}} = \epsilon^-_\mu \sigma^{\mu}_{\alpha\dot\alpha} =\sqrt{2}  \frac{\lambda_\alpha \tilde\mu_{\dot{\alpha}}}{[\tilde\lambda\tilde \mu]}, \qquad \epsilon^+_{\alpha\dot{\alpha}} = \epsilon^+_\mu \sigma^{\mu}_{\alpha\dot\alpha} = \sqrt{2} \frac{\mu_\alpha \tilde\lambda_{\dot{\alpha}}}{\lag\mu\lambda\rag},
 \eeq
 where $\mu,\tilde \mu$ are any reference spinors linearly independent of $\lambda$, $\tilde\lambda$, which is reflecting the gauge redundancy. Indeed, any transformation of $\tilde \mu$ can be written as~\cite{Cheung:2017pzi}
 \beq
 \tilde \mu \rightarrow \tilde \mu + z \tilde \mu + z^\prime \tilde \lambda,
 \eeq
 with $z$ and $z^\prime$ being complex numbers. Since the polarization vector is invariant under the scaling of $\tilde{\mu}$, it becomes
 \beq
 \epsilon^-_{\alpha \dot \alpha} \rightarrow  \epsilon^-_{\alpha \dot \alpha} + \frac{\sqrt{2}z^\prime}{ 1 +z} \frac{p_{\alpha\dot\alpha}}{[\tilde \lambda \tilde \mu]},
 \eeq
 which is just the residual gauge transformation preserving the condition $p_\mu \epsilon^\mu = 0$. 
 For example, we can choose
 \beq
\mu_\alpha =\left(
\begin{array}{cc}
c^*\\
 s
\end{array}\right),  \qquad \tilde \mu_{\dot \alpha} = \left(
\begin{array}{cc}
c\\
 s^*
 \end{array}\right),
\eeq
which correspond to the  the explicit formulae for polarization vectors:
\beq
\label{eq:pvml}
\begin{split}
 \epsilon^{+\mu}& = \frac{1}{\sqrt{2}}(0 ,  \cos\theta \cos\phi - i \sin\phi ,  \cos\theta \sin\phi + i \cos\phi  , - \sin\theta), \\
  \epsilon^{-\mu}& =\frac{1}{\sqrt{2}}(0 ,  \cos\theta \cos\phi + i \sin\phi ,  \cos\theta \sin\phi - i \cos\phi  , - \sin\theta) .\\
\end{split}
\eeq

The $n$-point helicity amplitudes are not continuous functions of the momenta, but rather the functions of the spinor-helicity variables:
\beq
\mathcal{M}_n(\lambda_a,\tilde{\lambda}_a)  \\
\eeq
where $a = 1, \cdots, n$ denotes  the particle indices and satisfy  the covariant  constraint:
\beq
\mathcal{M}_n(\omega_a \lambda_a,\omega_a^{-1} \tilde{\lambda}_a) = (\omega_a)^{-2h_a}\mathcal{M}_n(\lambda_a,\tilde{\lambda}_a) .
\eeq

 Let us consider the geometry of $n$-particle Momentum Conservation:
\beq
\begin{split}
\sum_{a=1}^n p_a^\mu &=0 \qquad \Leftrightarrow \qquad  \sum_a \lambda_{a\alpha} \tilde{\lambda}_{a\dot{\alpha}} = 0\\
\end{split}.
\eeq
We can think of this condition as imposing a constraint on the spinor vector space $\{\lambda_{a\alpha}\}$ or $\{\tilde\lambda_{a \dot \alpha}\}$. The condition can be fully explored by projecting into two linearly independent spinors. For $n = 3$, this is easy to solve, as we can choose either $\lambda_a$ or $\tilde{\lambda}_a$ as generic. For example, in the first case, we can project into the $|1\>$ subspace and the non-vanishing of $\<12\>, \<13\>$ implies the proportionality of $|2]$ and $|3]$. Similarly, in the other case,  non-vanishing of $[12], [13]$ implies the proportionality of $|2\>$ and $|3\>$. Finally we have two solutions:
\beq
\begin{split}
\text{Generic } \lambda & \Rightarrow  \tilde{\lambda}_1 = \langle23\rangle \tilde{\xi}, \qquad \tilde{\lambda}_2 =  \langle 31\rangle \tilde{\xi}, \qquad 
\tilde{\lambda}_3 = \langle12\rangle \tilde{\xi},\\
 \text{Generic } \tilde{\lambda}  &   \Rightarrow  \lambda_1 = [23] \xi, \qquad \lambda_2 =  [ 31]\xi, \qquad \lambda_3 = [12] \xi,
\end{split}
\eeq
with $\xi,\tilde{\xi}$ being some reference spinors. For the three-particle amplitudes,  we have
\beq
\label{eq:3ptamplitude}
\begin{split} 
\mathcal{M}_3 &= \left\{ 
\begin{array}{ll} 
 \left<12\right>^{h_3  - h_1 - h_2}  \left<23\right>^{h_1 - h_2 -h_3} \left<31\right>^{h_2 - h_3 - h_1}, \qquad  h_1 + h_2 + h_3 < 0\\
 \left[12\right]^{h_1+h_2 -h_3}  \left[23\right]^{h_2 + h_3 - h_1} \left[31\right]^{h_3 + h_1 - h_2}, \qquad  h_1 + h_2 + h_3 > 0\\
\end{array}
\right.
\end{split},
\eeq
where we also  demand that the amplitudes have a smooth limit in Minkowski signature where the brackets also go to zero.

\section{Massive spinor variables}
\label{app:msv}
The massive spinor variables are a bit more complicated than the massive case, where the little group is $\SU (2)$ instead of $\ISO (2)$. Consequently, the spinor variables carry the little group index $I$, which transform as the fundamental representation of $\SU (2)$:
\beq
p_{\alpha\dot{\alpha}} =   \lambda^{\ I}_\alpha \tilde{\lambda}_{I\dot{\alpha}} = |p^I\>[p_I|,
\eeq
which can be thought of as the sum of two rank-1 matrices $\lambda^{1}_\alpha \tilde{\lambda}_{1\dot{\alpha}}$,  $\lambda^{2}_\alpha \tilde{\lambda}_{2\dot{\alpha}}.$ For general complex momenta, the spinor variables  transform under  the fundamental representation of $W \in \SL (2,\mC)$:
\beq
\lambda^I \rightarrow  \lambda^J (W^{-1})^{\ I}_{J}, \qquad \tilde{\lambda}_I \rightarrow W_{I}^{\ J}\tilde{\lambda}_J.
\eeq
We adopt the following analytic continuation:
\beq
 \lambda^I(-p) =- \lambda^I(p), \qquad \tilde{\lambda}_I (-p) =  \tilde{\lambda}_I (p).
\eeq
The case of real momenta for positive energy can be obtained by imposing
\beq
(\lambda_\alpha^I)^* = \tilde{\lambda}_{I\dot{\alpha}},
\eeq
which implies
\beq
(\tilde\lambda^{I \dot\alpha})^* = -\lambda_I^\alpha.
\eeq
Note that the little group index $I$ is naturally raised or lowered, which is consistent with the fact that the  fundamental representation of $\SU (2)$ is self-conjugate.
We can regard the massive spinor variables as two matrices ($I$ is the column index in $\lambda$ and row index in $\tilde{\lambda}$).

Unlike the massless case, the on-shell condition $p^2 = m^2$ is not manifest in this decomposition, but  rather a constraint on the spinor variables:
\beq 
\det p =  \det \lambda \times \det \tilde{\lambda} = m^2.
\eeq
Without loss of  any generality, we can always choose
\beq
\text{det} \lambda = m ,\qquad \text{det} \tilde{\lambda} = m,
\eeq
where $m = \sqrt{E^2 - \vec p^2}$. With this convention, we have the following identities:
\beq
\lambda^{\ I}_\alpha \lambda_{\beta I} = m \,\varepsilon_{\alpha\beta}, \qquad \tilde\lambda^{I}_{\ \dot\alpha} \tilde{\lambda}_{I\dot\beta } = m \,\varepsilon_{\dot\alpha \dot\beta}, \qquad \lambda^{\alpha I }\lambda_{\alpha}^{\ J} = -m \,\varepsilon^{I J}, \qquad \tilde\lambda_{I \dot\alpha} \tilde{\lambda}_J^{\ \dot\alpha } = -m \,\varepsilon_{IJ}.
\eeq
By using the above formulae,  it is straightforward to obtain the spinor version of the Dirac equations:
\beq
p_{\alpha\dot{\alpha}} \tilde{\lambda}^{\dot{\alpha}I} = m \lambda_\alpha^I, \qquad p_{\alpha\dot{\alpha}} \lambda^{\alpha I} = - m \tilde{\lambda}^I_{\dot{\alpha}} .
\eeq
We can always expand the spinor variables in the bases of the little group space as
\beq
\lambda^{\ I}_\alpha = \lambda_\alpha \zeta^{-I} + \eta_\alpha \zeta^{+I},\qquad \tilde{\lambda}_{I\dot{\alpha}} = \tilde{\lambda}_{\dot{\alpha}} \zeta^{+}_I + \tilde{\eta}_{\dot{\alpha}} \zeta^{-}_I,
\eeq
where the eigenvectors  of the $z$-component spin operator with eigenvalues of $\pm \frac 12$  are given by
\beq
 \zeta^{+I}= \left(
\begin{array}{c}
1\\
0
\end{array}
\right), \qquad
\zeta^{-I}= \left(
\begin{array}{c}
0\\
1
\end{array}
\right),
\eeq
which satisfies
\beq
\zeta^{-I}\zeta^+_I = 1.
\eeq
By using the above identity,  the momentum matrix can be written in terms of the expansion spinors $\lambda(\tilde\lambda)$, $\eta (\tilde \eta)$, 
\beq
p_{\alpha \dot{\alpha}}= \lambda_\alpha \tilde{\lambda}_{\dot{\alpha}} - \eta_\alpha \tilde{\eta}_{\dot{\alpha}},
\eeq
and the on-shell condition becomes
\beq
\label{eq:onshell}
 \left<\lambda\eta\right> = m, \qquad [\tilde{\lambda}\tilde{\eta}] = m.
\eeq

Similar to the massless case, the real momentum matrix for positive energy can be parametrized as
\beq
\begin{split}
p_{\alpha\dot{\alpha}} &= \left(
\begin{array}{cc}
E - p \cos\theta& -p\sin\theta e^{-i\phi} \\
-p \sin\theta e^{i\phi}   &E + p \cos\theta
\end{array}
\right) = (E+p)\left(
\begin{array}{cc}
s s^*&  -c^* s^* \\
-cs   &c c^*
\end{array}
\right)+  (E-p)\left(
\begin{array}{cc}
c c^*& c^* s^* \\
cs   &s s^*
\end{array}
\right)
\end{split},
\eeq
with $p = |\vec{p}|$, and the on-shell condition is $E^2 - p^2 = m^2$. It is not difficult to see that the following choices of the spinor variables can do the job:
\beq
\label{eq:lambdaeta}
\begin{split}
\lambda_\alpha &= \sqrt{E + p} \left(
\begin{array}{c}
-s^*\\
c
\end{array}
\right) , \qquad \tilde{\lambda}_{\dot{\alpha}} = \lambda^*_\alpha =  \sqrt{E + p}\left(
\begin{array}{c}
-s\\
c^*
\end{array}
\right),\\
\eta_\alpha&=  \sqrt{E-p}  \left(
\begin{array}{c}
c^*\\
s
\end{array}
\right), \qquad \tilde{\eta}_{\dot{\alpha}} =- \eta^*_\alpha= -\sqrt{E-p}\left(
\begin{array}{c}
c\\
s^*
\end{array}
\right),
\end{split}
\eeq
which satisfy the on-shell condition in Eq.~(\ref{eq:onshell}) as can be verified directly.
In the high-energy limit, we have:
\beq
\lambda \sim \mO(\sqrt{E}) , \qquad \eta \sim  \mO(\frac{m}{\sqrt{E}}),
\eeq
and $\lambda$, $\tilde{\lambda}$ coincide with Eq.~(\ref{eq:lambda}) of the massless case. The polarization vectors for the spin-1 particles transform as symmetric rank-2 tensors under the $\SU (2)$ little group and they can be constructed by the tensor-product of $\lambda^I$ and $\tilde \lambda^I$:
\beq
\bm{\epsilon}_{\alpha \dot \alpha} \equiv\epsilon_{\alpha\dot{\alpha}}^{I_1 I_2}  =\frac{\sqrt{2}}{m}  \lambda_\alpha^{\{I_1}   \tilde\lambda_{\dot{\alpha}}^{I_2\}}  = \left\{ \begin{array}{ccc}\frac{\sqrt{2}}{m}  \lambda_\alpha^{I_1}\tilde{\lambda}_{\dot \alpha}^{I_2}, \qquad I_1 = I_2\\
\frac{1}{ m} \left(\lambda_\alpha^{I_1}\tilde{\lambda}_{\dot \alpha}^{I_2} +\lambda_\alpha^{I_2}\tilde{\lambda}_{\dot \alpha}^{I_1}  \right), \qquad I_1 \neq I_2
\end{array}
\right. .
\eeq
where we have adopted the same convention as~\cite{Durieux:2019eor}. 
The longitudinal and transverse polarization components can be extracted as the coefficients of 
$\zeta^{+ I_1}\zeta^{- I_2}$, $\zeta^{+ I_1}\zeta^{+ I_2}$, $\zeta^{- I_1}\zeta^{- I_2}$ and they are found to be
\beq
\begin{split}
\epsilon^0_{\alpha\dot{\alpha}}&=\epsilon^{-\frac12,\frac12}_{\alpha\dot{\alpha}}=\frac{ \lambda_\alpha \tilde{\lambda}_{\dot{\alpha}} + \eta_\alpha \tilde{\eta}_{\dot{\alpha}}}{m}, \quad \epsilon^-_{\alpha\dot{\alpha}} =\epsilon^{-\frac12,-\frac12}_{\alpha\dot{\alpha}}= \sqrt{2} \frac{\lambda_\alpha \tilde \eta_{\dot\alpha}}{m} , \quad
  \epsilon^+_{\alpha\dot{\alpha}} = =\epsilon^{\frac12,\frac12}_{\alpha\dot{\alpha}} = - \sqrt{2} \frac{\eta_\alpha \tilde \lambda_{\dot\alpha}}{m} \\
\end{split}.
\eeq
Plugging in the formulae in Eq.~(\ref{eq:lambdaeta}), we find the explicit formulae for  polarization vectors as follows:
\beq
\begin{split}
\epsilon^{0\mu}&= \frac1m(p,E_p \sin\theta \cos\phi , E_p \sin\theta \sin\phi,   E_p \cos\theta ), \\
 \epsilon^{\pm \mu}& = \frac{1}{\sqrt{2}} (0 ,  \cos\theta \cos\phi \mp i \sin \phi ,  \cos\theta \sin\phi \pm i \cos\phi  , - \sin\theta). \\
\end{split}
\eeq
The amplitudes for massive particles are functions of
 $\lambda^I, \tilde{\lambda}_I$,  which are  fully symmetric rank $2S$ tensors  for spin $S$ particles.

\section{Massive amplitudes as IR-deformation of the massless amplitudes}
\label{app:IRdef}
Under our parametrization of massive spinor variables, in the high energy limit, they approach the massless spinor-helicity variables as follows:
\beq
\begin{split}
&\lambda^{\ I}_\alpha \rightarrow \lambda_\alpha \zeta^{-I} +\mO(\frac{m}{\sqrt{E}}), \qquad \tilde{\lambda}_{I\dot{\alpha}} \rightarrow \tilde{\lambda}_{\dot{\alpha}} \zeta^{+}_I +\mO(\frac{m}{\sqrt{E}}).
\end{split}
\eeq
This may provide a way to think that the massive amplitudes as appropriate IR-deformation of the UV massless amplitudes, especially for particles with spins. For scalar particles, the transformations under the little group are trivial and they don't provide too much insight. We also confine ourselves to the on-shell three-particle amplitudes and leave the higher-pt amplitudes for the future. In addition, we are satisfied with considering about the relevant and marginal interactions. This corresponds to the total helicity of the 3-pt massless amplitudes smaller than or equal to one:
\beq
|h| =|h_1 + h_2 + h_3| \leq 1,
\eeq
as the mass dimensions of the associated couplings are given by
\beq
[g_h] = 1 - |h|.
\eeq
The first non-trivial example involves the fermion-fermion-scalar amplitudes and as shown in Eq.~(\ref{eq:3ptamplitude}), there are two kinds of marginal on-shell amplitudes:
\beq
\mM(1^{-\frac12},2^{-\frac12},3^0) = \<12\>, \qquad \mM(1^{+\frac12}, 2^{+\frac12}, 3^0) = [12].
\eeq
The IR deformation is straightforward:
\beq
\<12\>  \rightarrow \<12\> \zeta_1^{-I} \zeta_2^{-J} \rightarrow \<1^I2^J\> \equiv \<\1\2\>,
\eeq
where symmetrization is implicitly assumed. In Ref.~\cite{Arkani-Hamed:2017jhn}, this has been denoted as ``bolding''.  Similarly, for plus-helicity amplitude, we have
\beq
[12] \rightarrow [\1\2].
\eeq

The case of massive spin-1 particles is more interesting, as it is famously known that an extra degree of freedom is needed to go from massless to massive. We will pursue it by first noticing the following properties of the massless spinor variables for  total-plus 3-pt on-shell amplitudes:
\beq
\begin{split}
|1\rangle &=\frac{[23]}{[12]}|3\rangle = \frac{[23]}{[31]}|2\rangle 
\end{split},
\eeq
which can be expressed as
\beq
\label{eq:identity3pt}
\frac{[23]}{[31]}  = \frac{\langle 1\eta\rangle}{\langle 2 \eta\rangle } , \qquad  \frac{[12]}{[23]}  = \frac{\langle 3\eta\rangle}{\langle 1 \eta\rangle } , \qquad \frac{[31]}{[12]}  = \frac{\langle 2\eta\rangle}{\langle 3 \eta\rangle }, 
\eeq
where  $\eta$ is any reference spinor linearly independent with angle-bracket on-shell spinors $|1\>$, $| 2\>$, $|3\>$. The first  set of on-shell massless amplitudes we are interested in are
\beq
\mM(1^{0},2^{+1},3^0) = \frac{[12][23]}{[31]} , \qquad \mM(1^{0}, 2^{-1}, 3^0) =  \frac{\lag12\rag\lag23\rag}{\lag31\rag},
\eeq
and for simplicity, we have set the coupling constant to one. We can deform them to the massive amplitudes involving two vectors and one scalar by employing the relations in Eq.~(\ref{eq:identity3pt}). Naturally, we will choose $\eta$ as the expansion spinor variables of different  particles with on-shell constraint as in Eq.~(\ref{eq:onshell}). To be more specific, for the total-plus helicity amplitude, the procedure reads
\beq
 \frac{[12][23]}{[31]} \rightarrow   \frac{[12]\lrag{1\eta_2}}{m_2}  \rightarrow \sqrt{2} \frac{\lrag{\1\2} [\1\2]}{m_2}
\eeq
and similarly for the $CPT$ conjugate amplitude, we have
\beq
 \frac{\lag12\rag\lag23\rag}{\lag31\rag} \rightarrow   \frac{\lag12\rag[1\eta_2]}{m_2}  \rightarrow \sqrt{2}\frac{\lrag{\1\2} [\1\2]}{m_2}.
\eeq
Remarkably, the two helicity amplitudes are unified into one massive object, and the spurious poles in the massless amplitudes have turned into mass singularities for the massive amplitudes.

The final example we are presenting here is three vector on-shell amplitude, and up to permutation and $CPT$ conjugation, the relevant one is
\beq
\mM(1^{+1}, 2^{+1}, 3^{-1})  = \frac{ [12]^3}{[23][31]},
\eeq
and the deformation gives us two mass singularities:
\beq
 \frac{ [12]^3}{[23][31]} \rightarrow  \frac{ [12]\<3\eta_1\>\<3\eta_2\>}{m_1m_2}\rightarrow \sqrt{2}\frac{ [\1\2]\lrag{\3\1} \lrag{\3\2}}{m_1m_2}. 
\eeq
The systematic way to IR-deform the on-shell massless amplitudes to massive ones  has been explored recently in Ref.~\cite{Balkin:2021dko}.

\section{Constraints of the coupling constants from the Lagrangian}
\label{app:basrel}

In Ref. \cite{Cornwall:1974km}, the Lagrangian of the most general tree unitary theory with a finite spectrum of spin-0, 1/2 and 1 states is given as
\bea
\mL &=& - \frac{1}{4} \left( F_{\mu \nu}^a \right)^2 + \bar{q}_R i \left(\slashed{\partial} + i \slashed{A}_a \bar{R}^a \right) q_R + \bar{q}_L i \left(\slashed{\partial} + i \slashed{A}_a \bar{L}^a \right) q_L\nn\\
 && + \frac{1}{2} \left[ \left( \partial_\mu + iA_{a \mu} \bar{T}^a \right) \bar{\pi} \right]^2 - V(\bar{\pi}) - \bar{q}_L Y(\bar{\pi}) q_R - \bar{q}_R Y^\dagger (\bar{\pi}) q_L\nn\\
 && + \sum_{a=1}^{N_0} \frac{1}{2} \left( M_0^a \right)^2 \left( A_{a \mu} + \frac{1}{M_0^a} \partial_\mu \theta_a \right)^2,\label{eq:cfbl}
\eea
where $F_{\mu \nu}^a = \partial_\mu A^a_\nu - \partial_\nu A^a_\mu - f^{abc} A^b_\mu A^c_\nu$ is the field strength tensor for the gauge field $A^a_\mu$, and $\bar{T}$, $\bar{L}$ and $\bar{R}$ are the generators associated with the representation  for the scalars, left-handed and right-handed fermions, respectively. It is written in the following basis, which we will call the ``gauge basis'', of scalars and vector states:
\begin{itemize}
\item The generators of the broken group $G$, and consequently the basis of the vector bosons, are organized according to invariant subgroups of $G$. In particular, the structure constants $f^{abc}$  are in the  ``Cartesian'' basis such that $ f^{ade} f^{bde} =0 $ for $a \ne b$.  This tells us that if the index $a$ belongs to the invariant Abelian subgroup, the structure constants $f^{abc}$ vanish  for all the indices $b,c$.
\item The generators $\bar{T}^a$ associated with the scalars are block diagonalized so that each diagonal block corresponds to an irreducible representation of $G$. 
\end{itemize}
The $N_V$ vector fields are labelled by $a$, and the indices $1 \le a \le N_0$ are for the  invariant Abelian subgroups that have an explicit mass term, the explicit mass matrix being diagonalized to be $\left(M_0^2 \right)_{ab} = \delta_{ab} \left( M_0^a \right)^2$.   $\theta_a$ with $a$ running from 1 to $N_0$ are the redundant scalars in the St\ddu ckelberg formalism for the massive invariant Abelian vectors.
All the other physical or St$\ddot{\text u}$ckelberg  scalars are grouped by $\bar{\pi}_p = \pi_p + \eta_p$ with $p = N_0+1,\cdots,\bar{N}_S$, where the constants $\eta_p$ are the vacuum expectation values (vev's). $V( \bar{\pi})$ and $Y(\bar{\pi})$ are quartic and linear functions of $\bar{\pi}$, respectively.

The full set of constraints on the coupling constants in Eq. (\ref{eq:cfbl}) include not only the Lie algebra for the structure constants $f^{abc}$ and the generators $\bar{T}$, $\bar{L}$ and $\bar{R}$:
\bea
f^{abe} f^{cde} + f^{ace} f^{dbe} + f^{ade} f^{bce} = 0,& & [\bar{T}^a, \bar{T}^b] = i f^{abc} \bar{T}^c,\nn\\
\, [ \bar{L}^a, \bar{L}^b ]  = i f^{abc} \bar{L}^c, && [\bar{R}^a, \bar{R}^b] = i f^{abc} \bar{R}^c,\label{eq:lgbla}
\eea
but also the following conditions on the scalar interactions:
\bea
 V_{,p} ( \eta )  &=& 0, \label{eq:potvev}\\
\bar{T}^a \bar{\lambda}^b - \bar{T}^b \bar{\lambda}^a &=& if^{abc} \bar{\lambda}^c,\label{eq:invvev}\\
 V_{,p} ( \bar{\pi})  \left( \bar{T}^a \bar{\pi} \right)_p &=& 0,\label{eq:invpot}\\
\bar{L}^a Y (\bar{\pi} ) - Y (\bar{\pi} ) \bar{R}^a -  Y_{,p} (\bar{\pi} ) \left( \bar{T}^a \bar{\pi} \right)_p &=& 0, \label{eq:invfer}
\eea
where we have defined $N_V$ column vectors in the $\bar{N}_S$-scalars space:
\bea
\bar{\lambda}^a_p = \left\{ \begin{array}{ll}
\delta^{ap} M_0^a, & 1 \le p \le N_0 \\
i   \bar{T}^a_{pq} \eta_q, & N_0 < p \le \bar N_S
\end{array} \right. .
\eea
and the generators $\bar{T}$ in the $N_0$-scalars subspace are zero, i.e. $\bar{T}_{pq} = 0$ for $p,q = 1,\cdots, N_0$. Eq. (\ref{eq:potvev}) states that the scalar potential has a local minimum at $\bar{\pi}_p = \eta_p$, while Eqs. (\ref{eq:invvev}), (\ref{eq:invpot}) and (\ref{eq:invfer}) guarantee that the various coupling constants involved are invariant tensors of $G$. At the current stage, one can already diagonalize the fermion mass terms so that we have
\bea
 Y (\eta) = Y^\dagger (\eta) = \delta^{\feri \ferj} m_\feri.
\eea

The field variables in Eq. (\ref{eq:cfbl}) in general do not correspond to the mass eigenstates of the bosons that we use in the on-shell calculations. The Lagrangian corresponding to our parameterization of the coupling constants in the ``mass basis'' is given by:
\bea
\mL &\supset& -\frac{1}{4} \left(\partial_{\mu} W_{a\nu} - \partial_{\nu} W_{a\mu} \right)^2- C_{abc} \partial_{\nu} W_{a \mu} W_b^{ \mu} W_c^{ \nu} + \sum_{a=1}^{N_V} \frac{1}{2} m_a^2 W_{a \mu} W_a^{\mu} \nn\\
&& + \bar{\psi}_R i \left(\slashed{\partial} + i \slashed{W}_a  R^a \right) \psi_R + \bar{\psi}_L i \left(\slashed{\partial} + i \slashed{W}_a  L^a \right) \psi_L  - \sum_{\feri} m_\feri \left( \bar{\psi}_{\feri L}  \psi_{\feri R} + \bar{\psi}_{\feri R}  \psi_{\feri L} \right)\nn\\
&&+ \frac{1}{2} \partial_\mu \phi_i \partial^\mu \phi_i - \frac{1}{2} \sum_{i= 1}^{N_S} m_i^2  \phi_i^2 + F_{abi} W_{a \mu} W^{b \mu} \phi_i - G_{aij} W_{a \mu} \partial^\mu  \phi_i \phi_j \nn\\
&& - \frac{1}{6} P_{ijk} \phi_i \phi_j \phi_k - \frac{1}{24} K_{ijkl} \phi_i \phi_j \phi_k \phi_l - \left( \bar{\psi}_L H_i \psi_R + \bar{\psi}_R H^{ \dagger}_i \psi_L \right) \phi_i,\label{eq:lagmb}
\eea
where  we have $N_S$ physical massive scalar states. We have neglected all 4-pt interactions except for $\phi^4$, as the others will be determined by the 3-pt interactions by unitary and locality. The parameters in the two Lagrangians can be related by performing appropriate transformations for the scalar and vector fields. The vector boson mass matrix in Eq.~(\ref{eq:cfbl}) is given by
\bea
\left( M^2 \right)_{ab} =\sum_{p=N_0+1 }^{\bar N_S} \left( i \bar{T}^a \eta \right)_p  \left( i \bar{T}^b \eta \right)_p +  \sum_{a = 1}^{N_0}\left( M_0^a \right)^2 \delta_{ab}.
\eea
The above symmetric matrix can be diagonalized  by some orthogonal transformation $O_{ab}$ on the  vector fields:
\beq
(OM^2 O^{-1})_{ab} = M_a^2\delta_{ab}.
\eeq
After the transformation, we can work out the  linear mixing terms between the vector and scalar states and this will give us the  Goldstone boson fields as follows:
\bea
\sigma_a = \sum_{b=1}^{N_0} \frac{O_{ab} M_0^b}{m_a} \theta_b + \sum_{p=N_0+1}^{N_S} \frac{i   O_{ab} \bar{D}^b_{pq} \eta_q}{m_a} \pi_p = Q_{ap} \Pi_p,
\eea
where we have grouped the scalar fields $\theta_a, \pi_p$ into one array $\Pi_p$ with
\beq
\Pi_p = \theta_p, \qquad p = 1, \cdots N_0; \qquad \Pi_p = \pi_p, \qquad p = N_0+1, \cdots \bar{N}_S,
\eeq
and the rotation matrix is given by:
\bea
Q_{ap} = \left\{ \begin{array}{ll}
\frac{O_{ap} M_0^p}{m_a}, & 1 \le p \le N_0 \\
\frac{i   O_{ab} \bar{D}^b_{pq} \eta_q}{m_a}, & N_0 < p \le \bar{N}_S
\end{array} \right. .
\eea
One can see that $Q_{ap} = (O_{ab}/m_a ) \bar{\lambda}^b_p$, and $Q_{ap} Q_{bp} = \delta_{ab}$. We can treat $Q_{ap}$ as $N_V$ orthonormal vectors in the scalar space, expressed in the gauge basis. Then one can find another $N_S = \bar{N}_S- N_V$ orthonormal vectors $Q_{ip}$, such that $Q_{\tilde{i} q}$, which includes both $Q_{ap}$ and $Q_{ip}$, forms a new, complete orthonormal basis in the scalar space. The physical scalar bosons are then given by
\bea
\phi_i = Q_{ip} \Pi_p,
\eea
and together with the Goldstone bosons $\sigma_a$ they form a new scalar basis $\Phi_{\tilde{i}}$, which is related to the gauge basis by the rotation
\bea
\Phi_{\tilde{i}} = Q_{{\tilde{i}}q} \Pi_q,
\eea
where the square matrix $Q_{{\tilde{i}} q}$ is orthogonal. (One of course has the freedom to choose $Q_{ip}$ so that the mass matrix of the physical scalars are diagonalized.)

We have thus figured out the rotation matrices $O_{ab}$ and $Q_{\tilde{i} p}$ needed to transform the vector and scalar states to their mass basis. To arrive at Eq. (\ref{eq:lagmb}) where all Goldstone scalars are eliminated, we need to use the form invariance of Eq. (\ref{eq:cfbl}) under gauge transformations. The actual transformations used to relate field variables in Eqs. (\ref{eq:cfbl}) and (\ref{eq:lagmb}) are~\cite{Cornwall:1974km}
\bea
\theta_p  &\equiv&  \phi_i Q_{ip} + \sum_{a=1}^{N_V} m_a   Q_{ap} \sigma_a, \qquad \bar{\pi}_p = \left[e^{i \sigma \cdot \bar{T} }\right]_{pq}  \left(\eta_q + Q_{iq} \phi_i \right), \nn \\
A_{a \mu} &=& \left[e^{ \sigma \cdot f} \right]_{ab} O_{cb} W_{c \mu} - \left[ \frac{e^{ \sigma \cdot f}}{\sigma \cdot f} \right]_{ab} \partial_\mu \sigma_b, \nn \\
q_R &=& e^{i \sigma \cdot \bar{R} } \psi_R, \qquad q_L =e^{i \sigma \cdot \bar{L} } \psi_L,
\eea
where we have defined
\bea
\qquad \left( \sigma \cdot f \right)_{ab} \equiv \sigma^c O_{cd} f^{abd},
\eea
and for any generator $\bar{T}_\text{r}$ of representation $R_\text{r}$ in the gauge basis,
\bea
\sigma \cdot \bar{T}_\text{r} \equiv \sigma^a O_{ab} \bar{T}_\text{r}^b.
\eea
Notice that the fermion masses remain diagonalized as a consequence of Eq. (\ref{eq:invfer}). Upon the above transformations, the mass basis couplings in Eq. (\ref{eq:lagmb}) can be expressed in terms of the gauge basis couplings in Eq. (\ref{eq:cfbl}) as
\bea
C_{abc} &= &f^{a'b'c'} O_{aa'} O_{b b'} O_{c c'},\qquad R^a = O_{ab} \bar{R}^b,\qquad L^a = O_{ab} \bar{L}^b,\nn\\
G_{aij} &=&- iT^a_{ij},\qquad F_{abi} =  -\frac{i}{2} \left( m_a T^b_{ia} + m_b T^a_{ib} \right),\nn \\
H_i &=& Q_{ip} Y_{,p},
\eea
where the scalar generator $T^a_{ij}$ in the mass basis is given by
\bea
T^a_{\tilde{i} \tilde{j}} = O_{ab} Q_{\tilde{i} p} Q_{\tilde{j} q} \bar{T}^b_{p q}.
\eea

Now we can figure out the constraints corresponding to Eqs. (\ref{eq:lgbla}), (\ref{eq:invvev}) and (\ref{eq:invfer}) in the mass basis. The coupling matrices  $T,L, R$  will still satisfy the commutation relations with $C_{abc}$ as the structure constants:
\bea
C_{abe} C_{cde} + C_{ace} C_{dbe} + C_{ade} C_{bce} = 0,& & [T^a, T^b] = i C_{abc} T^c,\nn\\
\, [ L^a, L^b ]  = i C_{abc} L^c, && [R^a, R^b] = i C_{abc} R^c.
\eea
In addition, by using
\bea
O_{ab} Q_{\tilde{i} p} \bar{\lambda}^{b}_{p} = m_a \delta^{a\tilde{i}},
\eea
we can rewrite Eq. (\ref{eq:invvev}) in the mass basis as follows:
\bea
T^a_{ib} m_b - T^b_{ia} m_a = 0,\qquad T^a_{cb} m_b - T^b_{ca} m_a = iC_{abc} m_c,\label{eq:d3vor}
\eea
which leads to
\bea
 F_{abi} = -im_a T^b_{ia}, \qquad  T^a_{bc} = iC_{abc} \frac{m_a^2 - m_b^2 - m_c^2 }{2m_b m_c}.
\eea
To summarize, the generator $T^a_{\tilde{i} \tilde{j}}$ can be completely expressed in terms of $ C_{abc}$, $G_{aij}$, $F_{abi}$ and the gauge boson masses, as in Eq. (\ref{eq:brbos}). Similarly, Eq. (\ref{eq:invfer}) leads to Eqs. (\ref{eq:brfer0}) and (\ref{eq:brfer1}).

\bibliographystyle{utphys}
\bibliography{references,refadd}

\end{document}